\documentclass[a4paper,12pt]{article}
\usepackage{dcolumn}%
\usepackage{bm}%
\usepackage{color}
\usepackage{epsfig}
\usepackage{cite}
\usepackage{amsfonts}
\usepackage{amsmath}

\usepackage{graphicx}%
\usepackage{dcolumn}%
\usepackage{bm}%

\newcommand{\be}{\begin{equation}}
\newcommand{\ee}{\end{equation}}
\newcommand{\ba}{\begin{eqnarray}}
\newcommand{\ea}{\end{eqnarray}}

\begin{document}
	\title{A Fluctuation Theory of Topological Susceptibility} 
	\author{Bhupendra Nath Tiwari$^{a, b}$ \\
	\\
	$^a$ INFN-Laboratori Nazionali di Frascati,\\
	Via E. Fermi 40, 00044 Frascati, Rome,\\ ITALY \\
\\
	$^b$ University of Information Science and Technology,\\
	``St. Paul the Apostle", Partizanska Str. bb 6000 Ohrid,\\
	REPUBLIC OF MACEDONIA\\
}
\maketitle
\begin{abstract}	
We investigate the long-range statistical correlations, whereby discuss the nature of the undermining interacting/ noninteracting domains and associated phase transitions under variations of the quark mass and the mass scale that corresponds to renormalized pion masses and the dimensions of an ensemble of slab sub-volumes of an arbitrary simulated lattice. The purpose of this paper is to compute the system's stability and its phase structures when it's model parameters vary infinitesimally. In particular, we focus on the stability properties and phases of an arbitrary (2+1) flavor QCD configuration under fluctuations of its parameters. In order to investigate the nature of statistical and systematic errors, and the presence of noises in the system, we explore fluctuation theory equivalences of the slab sub-volume method of computing the topological susceptibility with its low energy ChPT counterpart. Hereby, we find that the ChPT configurations always correspond to a non-interacting statistical basis in the space of the quark mass and the mass scale that corresponds to renormalized pion masses. The second system as an ensemble of finite slab sub-volumes of a simulated lattice turns out to be generically interacting under fluctuations of the slab dimensions. However, it yields an ill-defined degenerate system in the infinitesimal limit of slab parameters. It is worth mentioning that implications of the intrinsic geometric analysis are well suited towards the modeling based understanding of the gluonic topological charge density fluctuations, quark mass dependence and long auto-correlation of the global topology. Finally, we discuss the stability properties of sub-volume simulated lattice improvements towards the understanding of QCD vacua, the behavior of UV divergences, finite-volume effects, statistical precision, simultaneous measurements, and associated quantum channel measurements.   
\end{abstract} 

{\bf Keywords}: (2+1)-Flavor QCD; Topological susceptibility; Chiral fermions; Statistical Fluctuations; Intrinsic geometry; Noise Instabilities.\\

\newpage

\section{Introduction}

Studying fluctuations of the topological susceptibility has been an important issue in lattice QCD \cite{sj, aabefhi, brty}. One of the main concern is the sensitivity of the topological charge and its relation to the violation of the chiral symmetry \cite{nf}. It is worth recalling that the quark mass dependence of the topological susceptibility largely arises from a quark sea or small quantum fluctuations that are suppressed by orders of $\hbar$, see \cite{bcnw} for physics at a fixed topology and \cite{abeg} for the full QCD with two flavored quark simulations. In this concern, the respective contribution arising from the discretization systematics to the topological susceptibility is usually large \cite{Aoki, dgp}. 

Lattice theory based simulations of the QCD \cite{lm, afn} becomes a challenging problem as the lattice becomes sufficiently finer. This is because the global topological charge turns out to be frozen along a numerical data set \cite{asw}, also see \cite{1} for developments based on the Monte Carlo simulations. Hereby, we explore the question of the topological susceptibility fluctuations as a function of the quark mass in a varied range of the mass renormalization scales.

In this paper, following the above difficulties, we study mass fluctuations of the topological susceptibility in the light of the ChPT formulation and slab sub-volume method \cite{2, 1, 3,4, Aoki}. To be precise, let $m=m_{ud}$ be the up and down quark mass and $M= M_{phy}$ be the physical mass scale corresponding to the renormalized mass of pions, then the ChPT model based \cite{1, 2,3,4, Aoki} topological susceptibility is given by

\begin{equation} \label{akoi}
\chi(m, M):= \frac{\Sigma}{2}\bigg[ m- \frac{3 \Sigma}{16 \pi^2 F^4_{phy}} m^2 \ln \bigg( \frac{2\Sigma}{F^2_{phy}} \frac{m}{M^2} \bigg) + \frac{4 \Sigma l}{F^4_{phy}}m^2 \bigg]
\end{equation}

As far as the renormalization is concerned, the standard understanding of QCD is that there is a physical mass scale, usually called $\Lambda_{QCD}$ that is independent of the quark bare masses, see \cite{lqcd} towards the static potential analysis of two flavor quarks. Instead of $\Lambda_{QCD}$, one could take this mass to be any QCD mass, such as the proton mass, or the parameter usually called $F_{\pi}$ in pion physics (see \cite{pp} towards an associated study in the realm of the holographic QCD) that does not vanish in the limit when the quark bare masses are taken to zero \cite{wittenp}. However, in a generic framework of the standard model, it is worth mentioning that the renormalized pion mass as this mass parameter does not work well because it vanishes in the limit when the quark bare masses vanish, see Muta \cite{Muta} for an introduction to QCD fundamentals. Following the above observations, we explore fluctuation properties of an ensemble of QCD vacua through the Riemannian geometric arguments \cite{rga}. Intrinsically, by considering the topological susceptibility as an embedding function \cite{rga, bntef} from the space of the quark mass $m$ and the physical mass scale $M$ that corresponds to a renormalized mass of pions to the set of real numbers, we analyze its stability structures and possible phases by computing the associated intrinsic geometric invariants. In this concern, there exists a thermodynamic geometric understanding of the evolutions, correlations and phase transitions, see \cite{fisica} in the realm of statistical mechanical applications, and \cite{klp} towards QCD phase transitions.

In Eqn.(\ref{manifold}), the mass scale $M$ related to cutoff is kept arbitrary in order to have renormalized pion masses. In the sense that there exist various statistical and systematic errors, and the presence of noises in the system, the pair $(m, M)$ fluctuate independently. Such an intrinsic consideration renders a Riemanian geometric understanding \cite{rga,bntef,fisica} of the statistical errors, and systematic errors with a variable mass scale in order to get a renormalized  mass of pions. It is worth mentioning that the mass of pions arises as a function of the quark masses $m_u $ and $ m_d$ at a particular cutoff. For example, in the $\overline{MS}$ scheme,  we have a mass independent subtraction, see \cite{ms} towards renormalization theory and the optimized QCD scaling. Namely, the up and down quark mass ratio $m_u/ m_d$ may be extracted from the pion mass $M_{\pi}$ by using the chiral symmetry. Further, the presence of various statistical fluctuations and noises in the system support that the mass scale $M$ and the quark mass $m$ vary independently. In particular, the limit either $m=f_1(M)$ or $M=f_2(m)$ allows one to examine the chiral symmetry restoration \cite{add1} and phase transition curves \cite{widom} as a specific submersion in the space of $\{ m, M \}$. Indeed, the optimal estimation of the quark masses $m_u$, $m_d$ and their statistical precision is the subject matter of an active research, see \cite{add1} towards recent investigations.

In order to analyze the system's stability and associated phase transitions through Widom curves \cite{widom} of an arbitrary QCD configuration, we investigate the topological susceptibility $\chi$ as the model embedding \cite{bntef} under fluctuations of the parameters $\{m, M\}$. Namely, by viewing the topological susceptibility as in the above Eqn.(\ref{akoi}) as the embedding function 

\begin{equation} \label{manifold}
\chi:\mathcal{M}_2: \rightarrow \mathbb{R}
\end{equation}

assigning in reality the pair $(m, M)$ to $\chi(m, M)$, we investigate the nature of stability properties on fluctuation surface $\mathcal{M}_2$, see section 2 below for an overview of the fluctuation theory. Hereby, we examine local and global phase structures of $(2+1)$ flavor lattice QCD configurations in the framework of the intrinsic geometry. Other issues of the interest include dynamics of the chiral fermions, chiral symmetry breaking, coarse grained sub-lattices of a simulated lattice and the optimized formation of chiral condensates \cite{5, 6, 7} as a definite combination of low energy constants such as the renormalized pion mass $M_{\pi}$ and physical values of the decay constant $F_{phy}$, also see \cite{8} for related computations at the next to the leading order.

In the light of effective field theories, we explore the notion of confining-deconfining phase transition and associated chiral phase transitions with nonzero quark mass effects. Ref. \cite{Pisarki} offers a combined nature of the chiral and deconfining phase transitions. A general hypothesis of effective models is to keep the relevant symmetries and consider the effects of all the terms that obey the underlying symmetry. In this concern, we provide statistical estimates and qualitative behavior. Our consideration includes lattice QCD fluctuations and their consequences for the QCD phenomenology \cite{nf}. As discussed in section 2 below, prior studies models include the sigma model, NJL model and Polyakov loop model. In such cases, the chosen effective potential is anticipated to capture the model features, while the lattice data are exploited to fit the model parameters. For example, QCD phase transitions are described by varying quark masses that support pure gauge deconfinement to the chiral limit. Concerning the $SU(N_c)$ gauge theory, recall that there exits a gauge field invariance under the center of $\mathbb{Z}_{N_c}$. In such cases, one can determine the finite temperature boundary conditions through the gauge equivalence. 

Moreover, the order parameter of an effective model can be constructed from the Polyakov loop where the mass scale is introduced via the eigenvalues of the Polyakov loop \cite{Pisarki}. In this concern, one may perform the analysis by considering the free energy. At varying temperatures, one can try to examine the properties of barrier formation. It is expected that such barriers quickly disappear as the temperature increases. Further, the order parameter can be defined as the expectation value of a Polyakov loop \cite{Pisarki, DumitruPisarki}. In this direction, QCD dynamics, phase transitions, domain walls and spinodal compositions are important topics for further research. It would be pertinent to study chiral transitions in the light of effective QCD theories. Recall that the massless QCD configuration has $SU(N_f)_L \times SU(N_f)_R$ flavor symmetry. A detailed analysis of the mass term is outlined in section 2 below. Physically, we see that the quarks arise as the fast degrees of freedom and mesons as the slow degree of freedom. Statistical analysis is found by the coarse-graining of Landau-Ginzburg effective potential. It is worth examining whether one can explicitly break the symmetry by introducing masses to quarks. As far as the pions are concerned, the pion decay constant $f_{\pi}$ is experimentally determined to be $93\ MeV$ by computing their weak decay amplitude. Hereby, a nonzero pion mass softly breaks the axial current \cite{nf}, whereby the true vacuum is shifted, however, in this case, we can redefine $f_{\pi}$ to translate the true vacuum at its experimental value.

In this case, we can fix all the parameters of the model to reproduce the vacuum features of mesons. The connection with the quark mass is given by the Gellmann Oakes Renner relation
\begin{equation}
m_{\pi}^2 f_{\pi}^2 = \frac{m_u+m_d}{2} \langle 0|\overline{u} u+\overline{d} d|0 \rangle
\end{equation}
This gives a connection between the pion mass $m_{\pi}$ and the quark mass $m_q$, and the $\sigma$ condensate and the chiral condensate \cite{Pisarki}. In short, $m_{\pi}$ depends on $m_q$ and the condensate of low lying fields. In a medium, in a finite temperature field theory \cite{Pisarki, DumitruPisarki}, we consider the expectation value $\langle \sigma \rangle (T)$ and examine the melting of the chiral condensate $\langle 0| \overline{\psi} \psi |0 \rangle$ in the high temperature limit. Here, given the $\pi$ and $\sigma$ self-interactions, the quark mass plays an important role in understanding the explicit breaking of the chiral symmetry as per the expectation value
\begin{equation}
\langle 0 | h \sigma |0 \rangle = \langle 0|(-m \overline{\psi} \psi) |0 \rangle,
\end{equation}
where $h \sigma$ is the explicit chiral symmetry breaking term. For example, in the light of $\sigma$ model of an $O(4)$ field, given the concerning effective Lagrangian, we can write down the associated partition function, whereby integrating over the fermions that are in a heat bath for chiral fields in the configuration, one arrives at an effective thermodynamic potential of the underlying field. Hereby, the effective potential in $\sigma$ direction shows a very different dynamics \cite{Petreczky2004a} concerning the phase transitions at the RHIC and that at the early universe cosmological scales. 

Hereby, the phase diagrams involve the analysis of thermodynamic potentials. Nonzero quark mass effects are discussed via the chiral condensate and renormalized Polyakov loop for a positive quark mass on a lattice \cite{petrov2006}. Phase diagrams are studied in determining the QCD point in the space quark masses, pion mass and temperature. Interesting questions arise concerning the order parameters and susceptibilities in $SU(3)$ gauge theories with massive fundamental quarks, chiral and deconfinement transitions. Does the temperature is the same for the restoration of the chiral and deconfinement symmetries? Following Karsch \cite{Karsch2002a, Karsch2002b}, near a chiral critical point, the chiral and deconfinement symmetry restoration temperatures differ by about $25$ MeV. Herewith, our analysis offer an intrinsic geometric perspective to succeptibility fluctuations, whereby the statistical aspects of QCD ensemble stability criteria.

For light quarks, it is known that the chiral symmetry restoration drives the deconfinement. This is supported by $\sigma$ model and generalized Landau-Ginzburg theory \cite{Mocsy2003}. On the other hand, for very heavy quarks, the deconfinement induces the chiral symmetry restoration in the realm of Polyakov loop model \cite{DumitruPisarki}. In these approaches, there are studies concerning lattice models of Wilson lines involving $SU(N)$ chiral spins \cite{RBC, Petreczky2004b, ChenDeTar1987}, order parameters in chiral and deconfinement scalar models \cite{Moscy2004}, quark-gluon-plasma and QCD phenomenology \cite{Meisinger2002}, Polyakov loop and its coupling to NJL model \cite{Fukushima2004}. For example, massive quarks explicitly break the symmetries with extra terms, whereby there are no true order parameters in the light of generalized Landau-Ginzburg theory \cite{Wilczek}. Having such interactions, Polyakov loop mass turns out to be related to lattice screening radius when one uses Gellmann-Oakes-Renner relation. Determination of the order parameters is anticipated through reliable lattice data \cite{Petreczky2004a}. 

Following the same, we wish constructing useful effective models that are capable to describe deconfining phase transitions via the Polyakov loop \cite{Pisarki, DumitruPisarki} and its role in QCD phenomenology. One interesting situation may arise by adding fermions in the background in the light of the matrix model. Notice that the chiral condensate can be considered as the order parameter to describe the chiral phase transitions. Our analysis shows that the quark mass even if it is taken to be small can render nontrivial consequences to QCD phase structures. As mentioned above, there are various approaches to address the physics of deconfinement and chiral phase transitions. Such mass effects are examined hereby in the light of the fluctuation theory. 

Fluctuation theory \cite{rupp1, tcb, Jencova, ourpaper, bntfr, bookbtg, bntngsb, book2018, fisica, aman, halfbpspaper, Levin, hri, ct, bntef, 
qg, vnqm, tafb, OMEGA, Weinhold1, Weinhold2, rupp2, Ruppeiner3, Ruppeiner4, McMillan, Simeoni, widom} offers various notions to understand statistical and quantum  properties of an ensemble of observations and instrumental technology towards a high precision  measurement. \`A priori, we have investigated the above perspectives in the light of configurations emerging from condensed matter systems, e.g.,  \cite{rupp1, rupp2, fisica}, electrical engineering \cite{bookbtg, bntngsb}, computer science and engineering \cite{book2018}, and various high energy configurations such as the flat information geometries \cite{ourpaper, aman} and counting entropy \cite{halfbpspaper} that it arises as a fluctuation theory of counting systems whose degeneracy is described as a selection problem. Hereby, we have examined statistical correlations and instabilities undermining the combinatorics of selecting $n$ boxes out $m= N^2$ boxes, where $n\le m$.  
Further, we have extended studies of fluctuations for the configurations whose embedding function are defined by the Shannon entropy, R\`enyi entropy, Tsallis entropy \cite{tcb, Jencova} and other higher order corrected entropies in the large charge limit of observations \cite{ourpaper, bntfr}. Following the above motivations of the fluctuation theory, in this paper, we apply it's notions to lattice QCD configurations \cite{ct}. 

In this concern, we discuss the stability properties of fluctuations via the topological susceptibility as an embedding map \cite{bntef}, whereby examine what are the behaviors of data samples as a distribution of observations. In particular, we compute fluctuation properties through the topological susceptibility corresponding to an ensemble of (2+1) flavor QCD configurations with chiral fermions in the space of the mass of quarks and the mass scale pertaining to renormalized pion masses. We focus on the limit of the ChPT formulation and slab sub-volume method of the topological susceptibility. Following the flow-time dependence of Yang-Mills gradient flow, we explore statistical and systematic fluctuation properties and analyze the critical behavior and associated geometric invariants in the space of a given pair of dimensions of an ensemble of slab sub-volumes of an arbitrary simulated lattice as the model parameters. 

In the light of system's global stability, we investigate the long-range statistical correlations, whereby discuss the nature of the undermining interacting/ noninteracting domains and concerning phase transitions under variations of the quark mass and the mass scale that corresponds to renormalized pion masses. We equally examine the associated analysis under fluctuations of the slab dimensions of an arbitrary simulated lattice. In this concern, it is worth noticing that the parameters, viz. the quark mass and pion mass scale remain statistically independent because of nontrivial condensate formations of the associated fields. Similarly, the end points of the time slices between which the correlator is summed-up are independent on the single configuration of an arbitrary simulated lattice. The overall evolution of such a system is determined by the QCD action. An overview of the same in the realm of the phenomenological and effective field theories, global color model, running quark current mass, nonperturbative dynamics, functional hadronisation, chiral constituent pions, pionic fluctuations, and an exact constituent pion mass at a small nonzero quark current mass is relegated to section 2 below.

Hereby, we explore the fluctuation theory equivalence of the slab sub-volume method of computing the topological susceptibility with its low energy ChPT counterpart. In particular, we find that the ChPT configuration always corresponds to a non-interacting statistical basis in the space of quark mass and the mass scale corresponding to renormalized pion masses. The second configuration as an finite slab sub-volumes of a given lattice turns out to be generically interacting under fluctuations of the slab dimensions. However, we find that it yields an ill-defined degenerate statistical system in the infinitesimal limit of the slab dimensions.

In the realms of QCD, we investigate the local and global correlation structures under fluctuations of model parameters with the topological susceptibility as the embedding function. First of all, we consider the mass scale $M$ that corresponds to a renormalized mass of pions and the quark mass $m$ as the parameters and the topological susceptibility $\chi$ as the model embedding function as in Eqn. (\ref{manifold}). Physically, given the quark mass $m$, the limiting configurations with the mass scale $M$ as a function of $m$ and vice-versa arise as special cases of the present consideration. Hitherto, it follows that the configurations either with $M=f_1(m)$ or $m=f_2(M)$ arise as a submersion from the space $\mathcal{M}_2$ of the parameters $\{m, M\}$ to the set of real numbers $\mathbb{R}$, where $f_1$ and $f_2$ are real valued functions. For example, see Eqn.(\ref{cim0}) for an explicit expression of $f_1$ as $M(m)$ when $\chi$ has a constant dependence on the quark mass $m$.
	
On the other hand, in the light of slab sub-volumes of an arbitrary simulated lattice \cite{Aoki}, we consider an intrinsic geometric examination of long range auto-correlations. In particular, we focus on an ensemble of lattices whose spacing remains very small in comparison to the inverse of the mass of pions $1/M_{\pi}$. Hitherto, notice that the slab dimensions $\{t_1, t_2 \}$ fluctuate independently because of the (i) existence of temporal noises in slab sub-volumes of a given simulated lattice and (ii) contributions due to non-zero mass $m_0$ of $\eta^{\prime}$ mesons. This consideration further supports one in examining the nature of fluctuations of an ensemble of slab sub-volumes in an arbitrary time slice $(1/m_0, T-1/m_0)$. This equally holds with or without the presence of statistical errors and systematic noises for an ensemble of non-trivial local correlators with the topological charge density as the model embedding. Hereby, under fluctuations of the slab dimensions $\{t_1, t_2 \}$, we offer an intrinsic geometric classification of the stability, correlations and phase transition curves for an ensemble of slab sub-volumes of an arbitrary simulated lattice. 
	
Physically, the slab parameters $\{ t_1, t_2 \}$ can be viewed as the end points of the time slices of a given correlator that is summed up. As for as the slab sub-volume analysis is concerned, it follows from the fact that the slab sub-volumes fluctuate within their error bars under statistical fluctuations of a given QCD vacuum. Such an error implies that the slab dimensions $\{ t_1, t_2 \}$  fluctuate independently. Hereby, for a given pair of slab dimensions $t_1$ and $t_2$ varying arbitrarily, we investigate the ensemble stability analysis \cite{fisica} of lattice QCD configurations through the derivatives of the slab averaged topological charge density correlator with respect to $t_1$ and $t_2$. To be precise, let $t_c$ be the cutoff time and $t_0$ be an arbitrary reference time, our proposal offers system's stability under parametric fluctuations of the topological charge $ \langle Q^2  \rangle$ in the interval $(t_0, t_0+ t_c)$. As far as the lattice QCD samples are concerned, a system gets approximated in a linear cut-off region \cite{Aoki} when the cutoff $t_c$ satisfies the inequality involving the mass $m_0$ of $\eta^{\prime}$ particles that arises as the first excited states, see the Eqn.(\ref{linear}) below for an overview.  	
	
As far as the high energy configurations are concerned, our analysis incorporates nontrivial contributions of $\eta^{\prime}$ mass fluctuations into the consideration. This follows from the fact that the quark mass dependence of the topological susceptibility largely arises from the quark sea, namely, as the quantum fluctuations that are suppressed by some orders of the $\hbar$. As a special case, the topological susceptibility reduces to a function of $t_2-t_1 $ in the limit of a small critical time. Physically, this holds as a special case of the noise free configuration that can equally be viewed as the large $N$ limit of the considered lattice. In addition, for a given ensemble of slab sub-volumes, we see that the $t_2-t_1$ dependence of the topological susceptibility arises as a submersion from the space $\mathcal{M}_2$ of the slab dimensions $\{t_1, t_2\}$ to the set of real numbers $\mathbb{R}$. The associated discussion of infinitesimal slab fluctuations is relegated to Section 5.

Our formulation offers the optimal evaluation of the topological susceptibility in the following ways. First of all, the dynamics of quarks is based on the M\"obius domain wall fermions \cite{9} that realizes an apt chiral symmetry as the residual mass, which can be kept at a small value, say for instance with an energy scale of a few MeV or smaller. As illustrated below, our results mildly depend on the lattice spacing in contrast to the overlap fermion models, see \cite{5} for an instance. It is worth emphasizing that our fluctuation theory analysis uses domain wall fermionic action for gauge ensemble generations \cite{22,23,24} in the Symanzik gauge. This allows to have the optimal sample fluctuations of lattice QCD configurations in its different topological sectors, which are classified over an ensemble of varied sub-volumes of the simulated lattice.

Secondly, our analysis is intrinsically based on an ensemble of improved sub-volumes lattice simulations encompassing a finite QCD correlation length at an inverse scale of the pion mass $M_{\pi}$. We extend the analysis for the case where the topological susceptibility calculation goes beyond mere the utilization of the topological charge of the theory. The utility of a lattice having an ensemble of sub-volumes is equally tested in the literature concerning overlap quark simulations \cite{5,6,7}, where the topological susceptibility has been evaluated from finite volume correlator of the $\eta^{\prime}$ mesons \cite{10, 11}. However, our investigation is based on correlations of a locally non-negative topological charge density as originally proposed by Bietenholz et al. \cite{12, 13, 14} and explored further by Shuryak and Verbaarschot \cite{15}, Forcrand et al. \cite{16} and Aoki et al. \cite{Aoki}. In this regard, our proposition optimally reduces the statistical noise as the slab dimensions $\{t_1, t_2\}$ vary in an extraction of the topological susceptibility. Following the fact that the  M\"obius domain wall based formulation entails the most recurrent fluctuations of the topological susceptibility than it's global topological charge based extraction on a given lattice \cite{Aoki}, we evaluate the local and global statistical correlations under variations of $\{t_1, t_2\}$.

Improvements arise further by considering the topological susceptibility fluctuations with respect to the product of the square of a renormalized mass of scattered pions and the physical value of their decay constant, e.g. $F_{\pi}$ as mentioned before, on an ensemble of varying sub-volumes of a simulated lattice. This follows by taking a suitable combination of the next to the leading order low energy constants that are independent of the underlying chiral logs and finite volume approximations. Hereby, our statistical analysis is well protected from the effects emanating from a sea of strange quarks, see \cite{Aoki} for an introduction. Therefore, for a given value of the quark mass and the mass scale that corresponds to renormalized pion masses, we offer the optimal estimates of the statistical correlations and fluctuations of the topological susceptibility at physically acceptable values of the model parameters. Herewith, we focus on the fluctuation stability conditions at each data point in slab sub-volumes of the simulated lattice.

In nutshell, we investigate the intrinsic structure of fluctuations of the topological susceptibility by invoking the role of Yang-Mills gradient flow analysis \cite{17, 18, 19}. This enables a refined understanding of QCD, namely, in achieving (i) an integer value of the topological susceptibility, (ii) a controlled analysis of the UV divergences, and (iii) an optimally reduced noise free statistical ensemble. As far as effects of the sea quarks are concerned, our proposition renders the optimal statistical precision in the computation of the topological susceptibility in the limit of fluctuations over an equilibrium ensemble. This supports the optimal evaluation of chiral condensate formation as well as the associated correlation length in the limits of respective chiral and continuum configurations. This shows an apt agreement with the ChPT predictions as in Eqn.(\ref{akoi}) towards the mass of $\eta^{\prime}$ mesons. It follows further that our analysis holds for short distance extractions of the $\eta^{\prime}$ mass following the statistical correlation properties of the topological charge density and topological susceptibility fluctuations. 

From the framework of M\"obius domain wall fermion ensemble generations, our perspective directions include the instrumentation designing techniques, detector technology and their optimal precision analysis in Symanzik gauge, see \cite{22,23,24} towards large scale simulations and chirally symmetric fermions. The implications of the intrinsic geometry are well suited towards the modeling based understanding of the gluonic topological charge density fluctuations, quark mass dependence and long auto-correlation of the global topology. Following the same, in the subsequent section, we discuss geometric properties of sub-volume simulated lattice improvements towards a refined understanding of fluctuations in QCD configurations, behavior of UV divergences and finite volume effects in the light of the statistical precision, simultaneous measurements and ensemble stability via intrinsic geometric invariants.

The rest of the paper is organized as follows. In section 2, we give a brief review of the lattice QCD models, topological susceptibility computation over an ensemble of slab sub-volumes of an arbitrary simulated lattice and its relation to the fluctuation theory. In section 3, we examine fluctuations of the topological susceptibility through the ChPT formula under variations of the quark mass and the mass scale corresponding to renormalized pion masses. In section 4, we analyze the nature of fluctuations of the topological susceptibility by considering an ensemble of slab-sub-volumes of an arbitrary simulated lattice. In section 5, in the framework of the slab method, we discuss fluctuation theory analysis of the two point correlation function over an ensemble of slab sub-volumes when their dimensions are separated by an infinitesimal interval. Following the same, we provide physical implications of the results concerning the ChPT formulation and slab sub-volume fluctuations in section 6. Namely, by considering the topological susceptibility as the model embedding, we give a qualitative discussion of the results when the associated model parameters are allowed to fluctuate. In addition, we provide physical implications of the slab sub-volume method and compare its results with the existing models such as the ChPT predictions, and associated ALPHA and JLQCD collaborations. Finally, in section 7, we conclude the paper with issues for future research and investigations.     

\section{Review of the Model}

In this section, we provide a brief review of the lattice QCD models as proposed by Aoki, Cossu, Fukaya, Hashimoto, and Kaneko \cite{Aoki} and its relation to the fluctuation theory \cite{rupp1, tcb, Jencova, ourpaper, bntfr, bookbtg, bntngsb, book2018, fisica, aman, halfbpspaper, bntef, Weinhold1, Weinhold2, rupp2, Ruppeiner3, Ruppeiner4, McMillan, Simeoni, widom}. First of all, in order to invoke the setup of the lattice gauge theory, one considers a domain wall based fermionic action for gauge ensemble generations \cite{22, 23, 24} in the Symanzik gauge. Thus, the associated Dirac operator is evaluated in a step by step smearing of the underlying gauge links. The concerned simulations can be performed in different lattice sizes with a given set of input parameters, see \cite{25} towards a high precision scaling. Generically speaking, one chooses a few approximate values of the strange quark mass and for each value of the strange quark mass, wherefore one chooses a few values for the masses of the up and down quarks in order to simulate a large lattice with the same sample parameters that govern the finite volume effects, see \cite{Aoki} and references therein. 

The molecular dynamic based numerical simulations of such a sample are performed by using software packages. In this concern, the IroIro++  \cite{24} is well-capable of controlling the chiral symmetry violation for M\"obius domain wall fermions with residual mass $\approx 1 MeV$ \cite{26}. In particle physics, the pion decay constant $F_{phy}$ in the low-energy effective action determining the strength of the chiral symmetry breaking can be approximated as $F_{phy} \approx 130 MeV$, see \cite{Yao} for review on particle physics containing the summary of tables and abbreviated average measured properties of the mesons, quarks, gauge bosons, leptons, and baryons.

The physical values of the model parameters yielding the renormalized mass of pions and the decay constant are estimated by pseudo-correlators in a particular combination of the locally smeared operators, see \cite{27} towards PoS lattice analysis of QCD. In the light of the topology fluctuations in an ensemble of slab sub volumes of a simulated lattice, let us recall the Clover construction \cite{28} that is based the gluonic definition of a topological charge density, where the YM gradient fluctuations smooth the gauge fields in a lattice of small step-sizes, e.g., $0.5 fm$. 

Given such a lattice QCD configuration, one generically performs numerical evaluations of the Yang-Mills gradient flow in the setup of the Wilson gauge theory action. This possesses an infinitesimal step-size that depends on the parameters of the original QCD configuration. By considering it's topological susceptibility as the embedding function, we comprehend the local and global structures undermining the topological charge density fluctuations and associated statistical correlation properties. Hitherto, for a given auto-correlation time, one estimates the statistical errors via the jack-knife procedure, see \cite{1} for an overview of the model. 
\subsection{The pionic physics}
In this subsection, we provide an overview of the pion in the light of Gell-Mann-Oakes-Rennner mass formula for Nambu-Goldstone bosons in the standard QCD configurations \cite{cg}. Namely, we discuss the relevance of the pion masses and their dependencies on the quark-antiquark condensate $\langle q \overline{q} \rangle$ and quark current masses. Concerning the constituent pions, we focus on the full response function of the underlying quark propagators. In the presence of a nontrivial quark current mass with its running behavior, this gives the equivalence of the Nambu-Goldstone and Cahill-Gunner mass formulae \cite{cg} via the global color model of QCD as discussed below. 
\subsubsection{QCD phenomenology}
First of all, properties of pions continue to find fundamental interests in theoretical, phenomenological and experimental studies of QCD. Let us begin by recalling that pion is a nearly massless Nambu-Goldstone boson whose properties are directly related to dynamical chiral symmetry breaking and the dynamics of the concerning quarks and gluons. There have been various interests in understanding the mass formula for pions \cite{2a,3a,4a}. In particular, we focus on the Gell-Mann-Oakes-Rennner mass formula \cite{1a} and its equivalent mass formulae. Here, we study pion mass equivalences in order to appreciate different quantum field theories.  

Further, there exist various mass formulae for the pions that are known to interplay between the constituent pion Bathe-Salpter equation and nonlinear Dyson-Schwinger equation for the constituent quarks, see \cite{2a} for concerning discussions. This involves the response of the constituent quark propagator to the fine tuning of the current mass of the quarks. In the quantum field theory, different concepts are being used to distinguish the constituent pions, and the exact or the full pion, whereby we focus on its relevant mass expression with a quark running current mass function $m(s)$. Notice that the full pion mass formula is quite an intricate matter. Thus, in order to understand QCD phenomenology and lattice QCD models, our consideration aims to include a running current mass in the light of fluctuation theory. 

Notice that Gell-Mann-Oakes-Rennner is generic as it arises in either the case of the full or the constituent pions whether we are examining the QCD or some appropriate scheme to QCD like the global color model as long as we preserve the dynamics of chiral symmetry breaking and its activation in the light of underlying quark gluon dynamics. Hereby, the difference between the full or constituent pions lies on the consideration of the effective action. Normally, these effective actions refers to the constituent hadrons. 

Our analysis is based on limiting effective action of the chiral constituent pions, constituent quark propagator, and the associated Euler Lagrange equations undermining the hadronized global color model effective action. Hereby, it is worth mentioning that fluctuations about the configuration with the minimum configurations of effective action introduce constituent mesons. In this concern, the double counting problem, dressing operators, constituent pion mass formula and its fluctuations including the quark running mass and vector response function are the important topics that are addressed below in the light of the present research. 
\subsubsection{The running quark current mass}

First of all, it is anticipated that there are certain perturbative expansions for the mass of almost Nambu-Goldstone pions involving small quark current masses. This is constructed terms of the underlying non-perturbative quark gluon dynamics in its chiral limit. The associated low energy pion mass is discussed in the light of chiral symmetry breaking, current algebra and PCAC, see \cite{1a} for an introduction. As discussed in the introduction, for some given up and down quark masses $\{m_u, m_d\}$, the limiting pion mass $m_{\pi}$ in the realm of QCD models simplifies as 

\begin{equation}
m_{\pi}^2 = \frac{\rho}{f_{\pi}^2}(m_u +m_d), 
\end{equation}
where $\rho$ is the condensate parameter of the model that is defined as the quark-antiquark correlator $\langle q \overline{q} \rangle$. This shows that the mass of the pions is not merely the function of quark masses but it depends on the condensate parameter $\rho=\langle q \overline{q} \rangle$. In particular, apart from  the other parameters such as the ordinary quark masses $m_u$ and $m_d$, it depends on the running quark current mass. Thus, in the light of our fluctuation theory analysis, when the running quark mass $m(s)$ in taken into the consideration, the model parameters $m$ and $M$ are treated independently of each other. In the case of the $SU(3)$ model of QCD, the condensate parameter $\rho$ is defined by the following overlap integral
\begin{equation} \label{oi}
\rho= 12 \int \frac{d^4 q}{(2 \pi)^4 }\sigma_s(q^2)
\end{equation}
Here, $f_{\pi}$ is the standard pion decay constant and $\sigma_s$ is the value of the quark propagator in its chiral limit that is defined as follows. Let $\sigma_v$ be the scalar part of the quark propagator and $m(s)$ be the running quark current mass, then the full quark propagator \cite{cg} is given by 
\begin{equation}
G(q, m)= -iq . \gamma \sigma_v(s, m)+ \sigma_s(s, m)
\end{equation}
Following the introduction of \cite{cg}, let $A(s, m)$ and $B(s,m)$ be the quark correlation functions arising from the underlying nonperturbative quark gluon dynamics \cite{9a, 13a}, then the limiting running chiral and scalar parts of the quark propagator \cite{cg} can be expressed as 
\begin{eqnarray} \label{ab}
\sigma_s(s, m)&=& \frac{B(s,m)+m(s)}{sA(s,m)^2+(B(s,m)+m(s))^2} ,\nonumber\\
\sigma_v(s, m)&=&\frac{A(s,m)}{sA(s,m)^2+(B(s,m)+m(s))^2}
\end{eqnarray}
Notice that the overlap integral as in Eqn.(\ref{oi}) diverges in QCD. This is because of the fact that for a given $N_c=3$ model with $\lambda= 12/(33-2N_f)$, we have
\begin{equation}
lim_{s \rightarrow \infty}\ B(s) \rightarrow 1/s \ln(\frac{s}{\Lambda^2})^{1-\lambda},
\end{equation}
where $\Lambda$ is the QCD scale parameter and $N_f$ is the number of flavors. This introduces an arbitrary integration constant $M$ in the determination of the pion mass $m$ and the parameter $\rho$. Given such an integration cutoff, as far as the physical values of $m$ and $\rho$ are concerned, they are quoted with reference to a cutoff momentum say of the order of a few $GeV$. Considering a definite volume $V$ in which the QCD configuration is realized, then at a given $\Lambda$, the set $\{m, \rho \}$ can be expressed in terms of the set $\{m, M\}$. 

Hereby, for a given QCD scale $\Lambda$, the parameters of the model $m$ and $M$ are governed nonperturbative quark gluon dynamics. Indeed, there exist various studies in the realm of pionic physics involving the Gell-Mann-Oakes-Rennner pion mass formulae where they are examined via some operator product expansions \cite{4a}, QCD sum rules \cite{6a, 7a} and their extensions to finite energy cases and Laplace sum rules \cite{8a} and others. 

\subsubsection{Nonperturbative low energy dynamics}
In this subsection, we discuss the global color model \cite{cg} in order to understand the low energy hadronic regime of QCD. Concerning the hadronization of QCD as a functional integral \cite{9}, the respective correlation function with its kernal involving the gluonic string structures in order to have the gauge invariance of the model. Following the same, the pion correlation function are defined as the connected part of the amplitude. 

Namely, in connection with the low energy regime of QCD, the non perturbative hadronic is captured in a functional integral. In this approach, the associated correlation functions are defined by a kernal function, wherefore the gauge invariance is obtained from the gluon string structure functions. Hereby, the pion correlation functions arise as the connected part of the amplitude. In the light of $\pi-\pi$ scattering, the pion mass is specified by the location of the pole in reference to the center of mass momentum of the Fourier transform of the amplitude. Notice that such an amplitude remains translationally invariant. This defines the QCD observables such as the pion upon imposition of mass shell conditions.

Theoretically, the correlation functions are computed through the generating functional of QCD \cite{cg}. In a low energy regime, the interactions of hadrons are extracted from effective actions corresponding to the hadronic states, whose parameters are designed by data fitting to experiments. In this case, the source terms of hadrons are produced by a hadronic generating functional. The generation of gluons are realized by the gluonic functional integrations involving an exact pure gluon propagator. In the computing of such propagator, the aforementioned global color model truncation is described by a quark gluon field theory action. Herewith, the local color symmetry of QCD arise via the nonquadratic terms of the gluon propagator. 

Hadronization involves the transformation of the bilocal mesons and diquark fields. The derived action involves fields and their derivatives. The mesons are extended states arise naturally from such bilocal fields \cite{cg}. Namely, in the light of the connected components of gluon propagator, the pion are described as a correlation function between two bilinear quark structures. Hereby, the bosonization is achieved by the change of variables in the above functional integral via the generalized Fierz transformations entailing the color, flavor and spin structures.

Following the fact that the quark gluon dynamics is dominated by fluctuations \cite{cg}, we apply the theory of fluctuations to study it's statistical correlations. However, it is known that the hadronic functional integral involving the local fields is almost perturbative in its nature \cite{cg}. In this approach, the optimal computing of the propagators of the gluons, quarks, mesons and baryons play an important role. Faddev equations determine the mass shell conditions using diquark propagators by considering them as quark quark pairs within baryons. Following the same, we discuss the concerned constituent hadrons using gluonic dressing of quarks as below.

\subsubsection{Functional hadronisation techniques}

Let us begin by recalling that the computation of correlation functions is realized by functional hadronisation techniques \cite{cg} in the center of mass coordinate system of pion. This involves the pion mass $m_{\pi}$ as an effective action parameter of the model at a given renormalization scale. Notice that this parameter is rather distinct from the observable pion mass as defined via a perturbative effective action governing the constituent pion.

The constituent hadrons arise as an expansion around the minimum of the effective action, satisfying the associated Euler Lagrange equations. The corresponding solutions satisfies nonvanishing bilocal fields and vanishing diquark fields that is merely the Dyson Schwinger equation \cite{cg} of the constituent quark propagator in the so called rainbow approximation. Gluonic dressing is generated by an additional term that generates a minimum solution with a nonvanishing bilocal fields forming the condensate $q \overline{q}$. 
Hereby, there exists a nontrivial possibility of the formation of diquark fields \cite{cg}. However, for configurations with the vanishing diquark fields \cite{cg}, there are no diquark or antidiquark condensates. 

Fluctuations of the action can be considered as the curvature of bilocal fields, namely, the inverse of the curvature generates mesons propagators having ladder type exchanges \cite{cg}. On the other hand, the functional hadronic integrals give nonladder type diagrams. Further, the inverse of the curvatures in the diquark sector give diquark propagators involving ladder type diagrams between the associated constituent quarks \cite{cg}. Additional complications arise when we consider generalized bosonization involving meson and diquark fields. Taking an account of a generalized Fierz transformation, the right color singlet $q\overline{q}$ correlations, color antitriplet qq correlations in a given color singlet baryon. In the rainbow or ladder approximation, the constituent states are described through an effective hadronic action \cite{cg} concerning the meson, where the observable hadronic states are obtained by the fully dressed constituent states. In general, the dressing merely produces a small shift in the values of the model parameters. However, the same doesn't hold for the nucleons \cite{cg}, that is, the is a large mass shift mainly arising from pions and mesons that provide appropriate dressing for nucleons. Such meson dressing to nucleonic correlations can be considered by a nonlinear functional integral \cite{cg} whose fluctuations are examined below under the variations if it parameters.

\subsubsection{ChPT as the chiral limit of constituent pions}

In the limit of vanishing quark current mass, the effective action possesses an extra global symmetry in the space of flavor of the quarks, see \cite{cg} for an overview. This is basically a  chiral symmetry with gauge group $U_L(N_F) \otimes U_R(N_F)$ arising from the realization
\begin{equation}
\overline{q} \gamma_{\mu} q =\overline{q}_R \gamma_{\mu} q_R + \overline{q}_L \gamma_{\mu} q_L,
\end{equation}
where we have $q_R= P_R q$, $q_L= P_L q$,  $\overline{q}_R= \overline{q} P_R$, and $\overline{q}_L= \overline{q} P_L$. It is evident that these two terms are individually invariant under the transformations $q_R= U_R q_R$, $q_L= U_L q_L$,  $\overline{q}_R= U_R^{\dagger}\overline{q}$, and $\overline{q}_L= U_LR^{\dagger} \overline{q}$. In the light of the aforementioned global color hadronization  model, the concerned Euler Lagrange equations yield degenerate solutions that are manifestly seen in terms of the constituent quark propagator. The associated degeneracy of the minimum configuration gives some fluctuations that have a nonzero mass. This corresponds to massive states of the Nambu-Goldstone nonlinear Dyson-Schwinger equation. Herewith, we observe the realization of Goldstone theorem \cite{gt} in the light of diquark confinement.

While passing from the hadronic functional integral of bilocal fields to that of the local fields, new variables that describe the degenerate vacuum manifold corresponding to the minimum solution are introduced by the angle variables in the resulting model. Hereby, the  Nambu-Goldstone term of the hadronization renders the action for the constituent pion as the ChPT effective action \cite{14a}. Notice that the aforementioned ChPT Lagrangian arises in the chiral limit of the constituent pions. The related coefficients of the model are explicitly obtainable in terms of the quark correlation functions $A$ and $B$ as depicted above in Eqn.(\ref{ab}). In the setting of the nonperturbative quark gluon dynamics, $A$ and $B$ are determined by the pure gluon propagator with ghosts as in the global color model of QCD. The concerned higher order terms describe $\pi-\pi$ scattering. Generically, by fixing the various model parameters as in the ChPT effective action, there have been various studies examining the dependence of the ChPT model on the pure gluon propagator \cite{16a}. The full consideration of the Nambu-Goldstone meson and nucleon coupling are dealt by the hadronization procedure.

Our analysis does not stop here, but it continues relating various QCD phenomenological models including the Nambu Jona Lasinio model \cite{njl}, ChPT model \cite{14a}, Cloudy Bag model or MIT model \cite{mit}, quantum meson coupling model \cite{nms} and soliton models \cite{soliton}. The exact pure gluonic correlation function can further be computed from lattice models as we have highlighted it in the light of the global color model. All these are the low lying limiting configurations that are used in describing the QCD phenomenology.      

\subsubsection{Pionic fluctuations}

The global color model provides solutions to various hadronic integral equations. In particular, the minimum configuration of the bilocal fields is equivalent to solving the Dyson-Schwinger equation of the constituent quark propagator. This determines the nonperturbative quark gluon dynamics. Namely, the vacuum equation of the global color model are computed in the rainbow approximation, viz. one can calculate the values of $A$ and $B$ are computed through the convolution of the running chiral and scalar parts of the quark propagator as in Eqn.(\ref{ab}) with the exact pure gluon propagator with ghosts in a Feynman like gauge or a Landau gauge. This incorporates the quark gluon coupling with a running current quark mass \cite{cg}. Further, specific fluctuations arise from the nonzero linear momentum states. 

In short, having known the quark propagator, one can obtain the gluon propagator in the effective global color model. This provides an implicit mass shell condition that has its solutions solely in the time like region. Namely, the quark correlation functions $A$ and $B$ as in Eqn.(\ref{ab}) corresponding to an absolute minimum of the underlying bosonised effective action, see \cite{9a} for an extended discussion. Notice that the loop momentum is localized in a space like region that mixes the metric, whereby the quark and gluon propagators are close to a real space like region. This shows that the quark and gluon propagators are less known in the time like region. 

Following the same, we study fluctuations of the QCD models in the space of their parameters. In particular, we focus on the fact that the quark gluon dynamics is highly nonperturbative that even the vanishing mass limit, one can have nonperturbative solutions with nonvanishing bilocal meson fields. Hereby, we offer the fluctuation theory description of the dynamical chiral symmetry breaking with $B$ as the vanishing mass dominant amplitude, see \cite{23} for an overview of the diagrammatic classification and spin trace projection in the realm of $SU(N_F)$ gauge theory. Indeed, the limiting configurations with vanishing mass are compatible with the Goldstone theorem. In general, the quark correlation functions $A$ and $B$ can be computed from the Dyson-Schwinger equation of the constituent quarks.

\subsubsection{An accurate constituent pion mass}
In order to determine an accurate expression for the mass of the constituent pion at a small nonzero running quark current mass $m(s)$, one needs to compute an analytic solution to the vertex equation of order $m$ with the constituent quark propagators that are nonpertubatively defined for quark gluon dynamics. To do so, we may consider Taylor expansion of the constituent quark propagators in $m(s)$. In this setting, we may discuss the large space like $UV$ region and small space like IR region separately. In the IR regime, there exists a dynamical enhancement for the quark current mass through the gluon dressing. This shows a strong response of the limiting chiral constituent quark propagator by fine tuning the associated current quark masses in the IR regime \cite{Higashijima, Elias}. This provides a nonperturbative understanding of QCD quark effects through the limiting chiral constituent quark running mass. Here, the gluon and constituent quark correlation functions and the form factor of the state together play an important role in solving the dominant amplitude. As far as the constituent hadrons are concerned, there is an enhancement of the quark current mass in the IR region that rises quickly with its current mass \cite{12a}. Notice that the pion mass $m_{\pi}$ remains small for a small $m$. 

Using the Lorentz covariance, we may perform the analysis in the rest frame with an equal quark mass.  By performing the Fourier transform of the amplitude, one can express the pion mass $m_{\pi}$ as a variational mass functional for a given pion form factor. Its near chiral limit is realized by minimizing the above mass functional that is essential first order in $m$. Notice that the change in the zero linear momentum frame of the amplitude from its limiting chiral value is of the second order in $m$. In this case, the amplitude may be replaced by the constituent quark correlation function in determining the $m_{\pi}^2$ in the leading order in $m$. This gives an analytic expression for the constituent Nambu-Goldstone boson mass square, see \cite{2a} for an equivalence involving the condensate parameter $\rho$ that is defined as an overlap integral as in Eqn.(\ref{oi}). In this concern, there are manifestly different pion mass formulae that are equivalent to Langfeld Kattner constituent pion mass formula, which arises as a power expansion of $m(s)$ in the chiral limit, see \cite{cg} for a detailed evaluation of low energy QCD hadronic properties. 

From the above pair of identities, this offers the role of the constituent quark correlation functions $\{A, B\}$ satisfying Eqn.(\ref{ab}), whereby the nonlinear Dyson-Schwinger equation supports in understanding the constituent pions. Hereby, the spectrum of fluctuations of the bilocal action is governed by the Euler Lagrange equation. In particular, it is worth emphasizing that the global color model hadronization allows to comprehending the constituent states through their mass. Namely, for the observable nucleons, it is known \cite{cg} that the pion dressing incorporating nonladder diagrams entails a nontrivial significance towards the understanding of low mass states, viz. the scalar part of the constituent quark correlation determine the mass of the constituent pion. Thus, our fluctuation theory based modeling of QCD is realized through the hadronic effective actions involving states of the quarks, diquarks, mesons and baryons.
\subsection{Slab Sub-volume Method}
In this subsection, we provide a brief review of the topological susceptibility computation over an ensemble of slab sub-volumes of an arbitrary simulated lattice \cite{Aoki}. In the light of lattice QCD simulations, the global topological charge $Q$ suffers from a long range auto-correlation time when the lattice spacing is small or $Q$ drifts very slowly. In this case, one exploits behavior of the topological susceptibility in a sub-volume $\Delta V$ of the whole lattice volume $V$. Let the pion mass be $M_{\pi}$, then the QCD correlation length is limited up to the scale of $1/M_{\pi}$, whereby the topological susceptibility fluctuations can be extracted from sub-volume $\Delta V$ whenever its size is larger than the inverse mass $1/M_{\pi}$. Such an ensemble of sub-volumes is an effectively uncorrelated sample where instantons and anti-instantons can move without a barrier among different topological sectors. Thus, in contrast to the global topological charge, an observable is expected to have a smaller auto-correlation time in such an ensemble of sub-volumes.

Consider that we have a long simulated lattice of volume $V$. Consider its cutting into sub-volumes $\Delta V$ such that its size is larger than $1/M_{\pi}$. Then, the statistical analysis is performed by considering the ratio $V/\Delta V$. In this case, we have an uncorrelated sample of the total $V/\Delta V$ sub-lattices. Notice that in each topological sectors of such a sample, we have no potential barrier, whereby the instantons and anti-instantons can freely enter and go out from such a sub-volume \cite{12}. In order for having a short auto-correlation time, it is known \cite{Aoki} that the slab method turns out to statistically efficient way towards the computation of the topological susceptibility. Given a sub-volume with dimensions $x= (x_0, \vec{x})$ and $y= (y_0, \vec{y})$, as a function of the cutoff time $t_c$, the topological charge $Q$ is obtained \cite{Aoki} as the summation of two point correlation functions of the charge density $q$ as

\begin{equation}
 \langle Q^2(t_c) \rangle := \int_{t_0}^{t_0+t_c} dx_0 \int_{t_0}^{t_0+t_c} dy_0 \int d^3x  \int d^3 y\ \langle q(x) q(y) \rangle,
\end{equation}

where $t_0$ is an arbitrary reference time with the above cutoff $t_c$. In the above definition, the volume integrals over $\vec{x}$ and $\vec{y}$ are taken in the whole spatial region, whereas the time integrals over $x_0$ and $y_0$ are taken in a fixed interval $[t_0, t_0+ t_c]$. As the topological charge $ \langle Q^2(t_c) \rangle$ always remains positive, this method is suggested \cite{Aoki, 7} to be statistically more stable than the JLQCD and TWQCD collaborations \cite{JLQCD, TWQCD}, also see \cite{5} for associated details.

Considering a large enough lattice, the expectation value $ \langle Q^2(t_c) \rangle$ can be approximated as a linear function of $t_c$. With the leading order finite volume corrections, it is known \cite{Aoki} that
$ \langle Q^2(t_c) \rangle$ can be expressed as the sum of $t_c/t$ fraction of the topological susceptibility $\chi_t$ of the entire lattice of volume $V$ and an $\eta^{\prime}$ mass dependent term that vanishes identically at $t_c= T$, whereby we have a global topology. Let $m_0$ be the mass of such an excitation, e.g., $\eta^{\prime}$ particle, then we have a linear dependence of the topological susceptibility $\chi_t$ on $t_c$ as long as it satisfies the inequality

\begin{equation} \label{linear}
\frac{1}{m_0} \ll t_c \ll t- \frac{1}{m_0}
\end{equation}

Following the above linearity condition as in Eqn.(\ref{linear}), the topological susceptibility as a function of the reference thicknesses $\{t_1, t_2\}$ can be extracted \cite{Aoki} as per the formula

\begin{equation} \label{slab}
\chi (t_1, t_2):= \frac{T}{V} \bigg( \frac{\langle Q^2_{slab} (t_1) \rangle - \langle Q^2_{slab} (t_2)\rangle}{t_1-t_2} \bigg)
\end{equation}

Note that in a numerical analysis with cutoff $t_c= T$, one typically chooses the slab dimensions $t_1, t_2$ in an interval $(l_0, T/2)$, where $l_0$ is of the order of a few $fm$. It is worth mentioning that the above slab method is less noise than previous simulations \cite{Aoki, 5,7} and has more frequent fluctuations than the global topological charge. Hereby, one is capable of taking care of long auto-correlations undermining the global topology via the transnational invariance of the charge density in a given lattice. As shown in the next section, the slab sub-volume method has interesting properties towards its validation in the light of the fluctuation theory.

\subsection{Fluctuation Theory: An Overview}

In this subsection, we provide a brief overview of the fluctuation theory. Motivated from the above observations concerning the topological susceptibility, we consider the analysis of calculating the underlying statistical correlations and phase transition curves in the case of two or more parameter configurations. In general, for a function of more than one variable, say there are $n$ parameters in the system, the concerning phase transitions are mediated via global Riemannian geometric invariants of the embedding function 

\begin{equation} \label{manifoldn}
\phi:\mathcal{M}_n: \rightarrow \mathbb{R}, 
\end{equation}

Hereby, depending upon the nature of correlations, the configuration may have the uniform and nonuniform phases in the space of parameters. Namely, this requires to compute the undermining correlation volume in the space of parameters, that is, by Ruppeiner \cite{rupp1} is proposed to be proportional to the invariant scalar curvature of the intrinsic manifold, which corresponds to a fluctuating statistical basis with $n$ parameters. 
To be precise, considering an $n$ dimensional Riemannian manifold $\mathcal{M}_n$ endowed with a nondegenerate intrinsic metric tensor $g$ defined as the Hessian matrix of the embedding map $\phi$ with a real valued assignment $(x_1, x_2, \ldots, x_n) \mapsto \phi (x_1, x_2, \ldots, x_n)$. Hereby, the Ruppeiner's proposition \cite{rupp1, tcb, Jencova, ourpaper, bntfr, bookbtg, bntngsb, book2018, fisica, aman, halfbpspaper, bntef, Weinhold1, Weinhold2, rupp2, Ruppeiner3, Ruppeiner4, McMillan, Simeoni, widom} results in to the following statistical versus intrinsic geometric equivalence

\begin{equation} \label{corr}
V_{corr} \sim R(x_1, x_2, \ldots, x_n)
\end{equation}

In the case of two variables $(x_1, x_2)$, that is, for the embedding function $\phi:\mathcal{M}_2 \rightarrow \mathbb{R}$, it follows \cite{rupp1, tcb, Jencova, ourpaper, bntfr, bookbtg, bntngsb, book2018, fisica, aman, halfbpspaper, bntef, Weinhold1, Weinhold2, rupp2, Ruppeiner3, Ruppeiner4, McMillan, Simeoni, widom} that the corresponding correlation area varies as the curvature $R(x_1, x_2)$ (see the below Eqn.(\ref{cur})), whose discontinuities give the nature of underlying phase transitions on the fluctuation surface $\mathcal{M}_2$ of the system with a pair of parameters $\{ x_1, x_2 \}$.

Mathematically, given an embedding $\phi:\mathcal{M}_2: \rightarrow \mathbb{R}$ as the real assignment $(x_1, x_2) \mapsto \phi (x_1, x_2)$, it's Hessian matrix $H$ is given by the following relation

\begin{equation} \label{metric}
g_{i, j}:= \frac{\partial^2}{\partial x^i \partial x^j} \phi (x_1, x_2), \ i, j= 1, 2  
\end{equation}

For a given pair of parameters $\{ x_1, x_2 \}$ of the considered statistical configuration, it can be viewed as a fluctuation vector $\vec{x}=(x_1, x_2) \in \mathcal{M}_2$. Subsequently, the determinant of the metric tensor $g_{ij}$ on the fluctuations surface $\mathcal{M}_2$  as depicted in Eqn.(\ref{metric}) reads as

\begin{equation} \label{D}
D(x_1, x_2):= \begin{vmatrix}
\phi_{11}  & \phi_{12} \\ 
\phi_{12}  & \phi_{22} \\ 
\end{vmatrix}
\end{equation}

Given the above metric tensor as in Eqn.(\ref{metric}), there are various approaches to calculate  the Riemann-Christoffel tensors $R_{ijkl}$ of an intrinsic manifold $(\mathcal{M}_n, g)$, where $i, j, k, l= 1, 2, \ldots, n$, see  \cite{rupp1, tcb, Jencova, ourpaper, bntfr, bookbtg, bntngsb, book2018, fisica, aman, halfbpspaper, bntef, Weinhold1, Weinhold2, rupp2, Ruppeiner3, Ruppeiner4, McMillan, Simeoni, widom} for an instance.
Hereby, for the case of the intrinsic surface $(\mathcal{M}_2, g)$ fluctuations with its local coordinates $\{ x_1, x_2 \}$, it follows \cite{rupp1, tcb, Jencova, ourpaper, bntfr, bookbtg, bntngsb, book2018, fisica, aman, halfbpspaper, bntef, Weinhold1, Weinhold2, rupp2, Ruppeiner3, Ruppeiner4, McMillan, Simeoni, widom} that we can express the corresponding Riemann-Christoffel tensor $R_{1212}$ as the following determinant

\begin{equation} \label{N}
N(x_1, x_2):= \begin{vmatrix}
\phi_{11}  & \phi_{12}  & \phi_{22}  \\ 
\phi_{111}  & \phi_{112}  & \phi_{122}  \\ 
\phi_{112}  & \phi_{122}  & \phi_{222} 
\end{vmatrix}
\end{equation}

Herewith, the Ruppeiner scalar curvature \cite{rupp1, tcb, Jencova, ourpaper, bntfr, bookbtg, bntngsb, book2018, fisica, aman, halfbpspaper, bntef, Weinhold1, Weinhold2, rupp2, Ruppeiner3, Ruppeiner4, McMillan, Simeoni, widom} of the intrinsic surface $(\mathcal{M}_2, g)$ is defined as the ratio

\begin{equation} \label{cur}
R(x_1, x_2)= \frac{k_B}{2} \frac{N}{D^2},  
\end{equation}

where $k_B$ is the Boltzmann's constant and the numerator $N$ and the denominator $D$ of the scalar curvature $R(x_1, x_2)$ read as in Eqns.(\ref{D}, \ref{N}) respectively. This corresponds to the correlation area of the  statistical configuration under fluctuation of its parameters $\{ x_1, x_2 \}$.

Hitherto, given the embedding function $\phi: \mathcal{M}_2 \rightarrow \mathbb{R}$, the correlation length $l(x_1, x_2)$ of the statistical configuration - - when it is viewed as the above intrinsic manifold $\mathcal{M}_2$ with the local coordinates $\{x_1, x_2\}$ and endowed with the aforementioned metric tensor $g$  as in Eqn.(\ref{metric}) - - reads \cite{rupp1, tcb, Jencova, ourpaper, bntfr, bookbtg, bntngsb, book2018, fisica, aman, halfbpspaper, bntef, Weinhold1, Weinhold2, rupp2, Ruppeiner3, Ruppeiner4, McMillan, Simeoni, widom} as the square root of the above Ruppeiner's scalar curvature $R(x_1, x_2)$. In other words, it reads as 

\begin{equation} 
l(x_1, x_2) \sim \sqrt{R(x_1, x_2)}
\end{equation}

In this framework, there are no correlations in the system if the corresponding correlation length vanishes identically, that is, we have $l(x_1, x_2)= 0$. In other words, such a configuration corresponds to a non interacting statistical basis. Geometrically, we have an embedding function $\phi (x_1, x_2)$ such that $N(x_1, x_2)=0$. On the other hand, there exist phase transitions in the statistical system whenever the underlying correlation length $l(x_1, x_2)$ diverges, i.e., there exists an embedding function $\phi (x_1, x_2)$ such that we have $D(x_1, x_2)=0$ on the surface $(\mathcal{M}_2, g)$ under fluctuations of its model parameters $\{ x_1, x_2 \}$ provided that $N(x_1, x_2) \ne 0$. Such a curve $D(x_1, x_2)=0$ is termed as the degeneration curve where one of the parameters is defined in terms of the other, that is, algebraically, we have either $x_2 \equiv x_2(x_1)$ or $x_1 \equiv x_1(x_2)$. For the above systems, such an assignment corresponds to either a non-interacting statistical basis when $N(x_1, x_2)= 0$ or the system undergoes certain phase transitions when we have $\Delta(x_1, x_2)= 0$ in the space of systems parameters. Physically, in the first case, the system remains non-interacting, while the second case corresponds to a highly interacting statistical basis.

In the light of the intrinsic scalar curvature $R(x_1, x_2)$ as depicted above in the Eqn.(\ref{cur}), we examine its denominator that arises as the determinant square of the metric tensor $g$, namely, we explore the question whether it induces a nondegenerate metric structure on the fluctuation surface $(\mathcal{M}_2, g)$ or not? The concerned analysis is relegated to the next sections. Namely, by considering the embedding as the topological susceptibility, we wish to emphasize on the stability properties of QCD configurations under fluctuations of their parameters. In section 3, we explore the fluctuation theory analysis in the space of the mass of quarks and the mass scale that corresponds to renormalized pion masses with the topological susceptibility as in Eqn.(\ref{akoi}) as the model embedding function. In section 4, by considering the topological susceptibility as the embedding function as above in the Eqn.(\ref{slab}), we offer an intrinsic geometric understanding of fluctuations over the dimensions of an ensemble of slab sub-volumes of an arbitrary simulated lattice. The case of infinitesimal slab fluctuations is discussed in section 6, below.

\section{Mass Dependent Fluctuations}

In this section, we investigate the local and global correlations as the response to QCD fluctuations. In order to do so, we consider the topological susceptibility as the model embedding function from the space of model parameters to the set of real numbers. As mentioned before, such considerations have appeared in condensed matter physics \cite{bookbtg, bntngsb, book2018, klp, fisica, Weinhold1, Weinhold2, rupp2, Ruppeiner3, Ruppeiner4, McMillan, Simeoni, widom}, high energy physics \cite{bntfr}, QCD \cite{bct1, bct2, bct3}, string theory configurations \cite{ourpaper}, understanding of black holes, black branes \cite{aman, halfbpspaper, Levin, hri, ct, bntef} and other systems \cite{rupp1, tcb, Jencova, qg, vnqm, tafb, OMEGA}. It is worth mentioning that an apt determination of the stability of a system and its phases in which it lies \cite{widom} are important concerns in fundamental physics. Thus, we address the issue of the system’s stability and phase transitions by considering the topological susceptibility as the model embedding function. It is defined from the space of the quark mass $m$ and the mass scale $M$ corresponding to renormalized masses of pions to the set of real numbers, see \cite{bntef} for an introduction. 

Following the above setup, we study fluctuations of the topological susceptibility through the ChPT formula as in Eqn.(\ref{akoi}) under fluctuations of the up/ down quark mass $m$, and the mass scale $M$ that corresponds to re normalized pion masses.
In order to simplify the subsequent computation, let us define the model constants as per the followings

\begin{eqnarray} \label{notations}
a &=& \frac{\Sigma}{2},  \nonumber \newline \\
b &=& \frac{3 \Sigma^2}{32 \pi^2 F^4_{phy}},  \nonumber \newline \\
c &=& \frac{2 \Sigma}{F^2_{phy}},  \nonumber \newline \\
d &=& \frac{2 \Sigma^2 l}{F^4_{phy}}
\end{eqnarray}

With the notations as in Eqn.(\ref{notations}), the ChPT formula depicted in Eqn.(\ref{akoi}) reads as

\begin{equation} \label{akoi1}
\chi(m, M) = a m + \tilde{b} m^2 - b m^2 \ln m + b m^2 \ln M,
\end{equation}

where the coefficient $\tilde{b}$ is defined in terms of the model constants $\{ b, c , d \}$ as

\begin{equation}
\tilde{b} = d - b \ln c
\end{equation}

The flow components as the first derivatives of the topological susceptibility $\chi$ are 

\begin{eqnarray} \label{flow}
\chi_m &=& a + (2 \tilde{b} - b)m - 2 b m \ln m + 2 b m \ln M, \nonumber \newline \\
\chi_M &=& - \frac{b m^2}{M^2}
\end{eqnarray}

Flow analysis is performed by the  solving the equations $\chi_m= 0= \chi_M$. The equation $\chi_m=0$ implies either a vanishing quark mass $m=0$ or a large physical mass scale $M= \infty$. From the vanishing of the component $\chi_m=0$, we see that the concerned physical mass scale $M$ that corresponds to re normalized pion masses satisfies 

\begin{equation}  \label{cim0}
M(m)= \bigg(\exp{\frac{b-2 \tilde{b}}{2 b}} \bigg) \frac{m}{\exp{\frac{a}{2bm}}}
\end{equation}

In the limit of vanishing quark mass $m \rightarrow 0$, it follows that the physical mass scale vanishes, as well, that is, we have $M \rightarrow 0$. On the other hand, in the limit of $M= \infty$, the quark mass $m$ as a function of the physical mass scale $M$ satisfies the following nonlinear equation 

\begin{equation}
a+ \beta m - 2b m \ln m =0,
\end{equation}

where the coefficient $\beta$ depends on the physical mass scale $M$ as per the identification

\begin{equation}
\beta (M)= 2 \tilde{b}- b- 2b \ln M
\end{equation}

In the limit of a small quark mass, viz. $m \approx \ln m$, it follows that there exists a pair of values of $m$ that correspondingly reads as 

\begin{equation}
m_{\pm} = \frac{1}{2 b} \bigg( \beta \pm \sqrt{\beta^2+ 8ab} \bigg)
\end{equation}

Under fluctuations of the model parameters $\{ m, M\}$, the heat capacities are given by 

\begin{eqnarray} \label{flowcapacities}
\chi_{mm} &=& (2 \tilde{b} - 3b)+ 2 b \ln \frac{M}{m}, \nonumber \newline \\
\chi_{MM} &=& - \frac{b m^2}{M^2}
\end{eqnarray}

The local correlation between the quark mass and the physical mass scale $M$ is given by

\begin{equation} \label{lcorq}
\chi_{mM} = \frac{2 b m}{M}
\end{equation}

Under fluctuations of the quark mass $m$ and the mass scale $M$ that corresponds to renormalized pion masses, the Hessian matrix $H$ associated with the topological susceptibility $\chi$ as in Eqn.(\ref{akoi1}) as the embedding $\chi: \mathcal{M}_2 \rightarrow \mathbb{R}$ is given by

\begin{equation} \label{fm}
H:= \begin{bmatrix} \chi_{mm} & \chi_{mM} \\ \chi_{mM} & \chi_{MM} \end{bmatrix}
\end{equation}

Herewith, the determinant of the above Hessian matrix $H$ defined as

\begin{equation} \label{detm0}
\Delta(m, M):= \chi_{mm} \chi_{MM} -\chi_{mM}^2
\end{equation}

reads as per the following expression
\begin{equation}\label{detm}
\Delta= \frac{1}{M^2}\bigg[ 3 b^2 m - 2b \tilde{b} m- 4b^2 m^2 - \frac{2b^2m}{M^2} \ln \frac{M}{m} \bigg]
\end{equation}

Therefore, for a finite physical mass scale $M\ne \infty$, the determinant $\Delta$ vanishes identically whenever the pair $\{ m, M \}$ satisfy the following nonlinear equation 

\begin{equation}
\beta_0 -\beta_1 m + \beta_2 \frac{m}{M^2} \ln \frac{M}{m} =0,
\end{equation}
where the coefficients $\{ \beta_0, \beta_1, \beta_2 \}$ are given by
 
\begin{eqnarray}
\beta_0 &=& 3 b^2 - 2b \tilde{b}, \nonumber \newline \\
\beta_1 &=& 4b^2, \nonumber \newline \\
\beta_2 &=& 2b^2
\end{eqnarray}

Now in the limit of $m=0$, and $M=0$, from Eqn.(\ref{detm}), we see that the  determinant approaches to a large positive value, viz. we have

\begin{equation}\label{detml}
\lim_{m,\ M \rightarrow 0} \Delta(m, M) \rightarrow \infty
\end{equation}

Therefore, in the small limit of $m$ and $M$, there exist pathologies in this model. Under fluctuations of the parameters $\{ m, M \}$, the global stability of the system is determined via the signature of principal minors $\{ p_1, p_2 \}$. Here, the first principal minor $p_1:= \chi_{MM}$ takes a negative value for $b>0$. Namely, in the limit of small quark mass $m$ and the mass scale $M$ that corresponds to renormalized pion masses, from Eqn.(\ref{flowcapacities}), it follows that we have the following limiting value of the first principal minor

\begin{equation}\label{pml}
\lim_{m,\ M \rightarrow 0} p_1(m, M) \rightarrow -b
\end{equation}

This shows that the system has instabilities at the origin in the space of the parameters $\{ m, M \}$, whereby one finds an intrinsic geometric origin of the production of new particles. Mathematically, the Hessian matrix $H$ as in Eqn.(\ref{fm}) defines a nondegenerate metric tensor $g:= (H_{in})_{2 \times 2}$ on the surface of fluctuations $\mathcal{M}_2$ whenever we have $p_2:= \Delta(m, M)>0$ and $p_1:= \chi_{mm}>0$. In order to compute the global correlation length \cite{rupp1, tcb, Jencova, ourpaper, bntfr, bookbtg, bntngsb, book2018, fisica, aman, halfbpspaper, bntef, Weinhold1, Weinhold2, rupp2, Ruppeiner3, Ruppeiner4, McMillan, Simeoni, widom} of the underlying statistical configuration, we need to evaluate the scalar curvature of the fluctuation surface $\mathcal{M}_2$ as in Eqn.(\ref{cur}). First of all, this requires evaluating the following determinant 
 
 \begin{equation} \label{nmatrix}
 N:= \begin{vmatrix}
 \chi_{mm}  & \chi_{mM}  & \chi_{MM}  \\ 
 \chi_{mmm}  & \chi_{mmM}  & \chi_{mMM}  \\ 
 \chi_{mmM}  & \chi_{mMM}  & \chi_{MMM} 
 \end{vmatrix}
 \end{equation}
 
 In this case, in order to calculate the global correlation length of the system with the above surface with $\chi(m, M)$ as in Eqn.(\ref{akoi1}) as the embedding function on the fluctuation surface $\mathcal{M}_2$, we find the below third derivatives 
 
 \begin{eqnarray} \label{thirdderi}
 \chi_{mmm} &=& -\frac{2 b}{m}, \hspace*{0.9 cm} \chi_{mmM}= \frac{2b}{M}, \nonumber \newline \\ \chi_{mMM}&=& -\frac{2 bm}{M^2},  \hspace*{0.3 cm} \chi_{MMM} = -\frac{2 bm^2}{M^3}
 \end{eqnarray}
 
Substituting the above expressions of the second and third derivatives of $\{\chi_{ij},\ \chi_{ijk}\ | \ i, j, k= m, M \}$ as in Eqns(\ref{flowcapacities}, \ref{lcorq}, \ref{thirdderi}), we see that the above determinant $N$ vanishes identically for all values of the model parameters $\{m, M\}$. This implies from Eqn.(\ref{cur}) that the Ruppeiner scalar curvature equally vanishes for all values of $\{m, M\}$. From Eqn.(\ref{corr}), following the definition of the correlation area of a statistical configuration as the Ruppeiner's scalar curvature \cite{rupp1, tcb, Jencova, ourpaper, bntfr, bookbtg, bntngsb, book2018, fisica, aman, halfbpspaper, bntef, Weinhold1, Weinhold2, rupp2, Ruppeiner3, Ruppeiner4, McMillan, Simeoni, widom} of the fluctuation surface $\mathcal{M}_2$, the fluctuating system having the ChPT formula of $\chi$ as its embedding function yields a noninteracting statistical basis.

Physically, it is worth anticipating that there are no phase transitions or global pathologies when the ChPT formula  of the topological susceptibility \cite{Aoki} is taken as the embedding function on the surface of the quark mass $m$ and the mass scale $M$ corresponding to renormalized pion masses. Hereby, we find that the ChPT formula  based topological susceptibility leads to a non-interacting statistical basis. It follows further that such a system possesses a vanishing statistical correlation length, that is, the limit of a large mass of pions, viz. $M_{\pi}\rightarrow \infty$.
 
\section{Slab Thickness Fluctuations}

In this section, we consider an intrinsic geometric investigation of the long range auto-correlations. In particular, we focus on an ensemble of lattices whose spacing remains very small in comparison to the inverse of the mass $1/M_{\pi}$ of pions. Further, it is known \cite{Aoki} that the QCD correlation length is equally inversely proportional to the mass of pions. Consequently, we find that fluctuation samples with the topological susceptibility as the embedding function \cite{bntef} are notable in the sub-volumes of dimensions larger than the QCD correlation length. It is worth mentioning that our analysis offers the optimal characterization of an arbitrary simulated of QCD lattice of the topological susceptibility as an observable of the auto-correlation time. 

Notice that the slab dimensions $\{t_1, t_2 \}$ - - as the end points of the time slices of an arbitrary correlator - - fluctuate independently because of (i) an existence of the temporal noises in slab sub-volumes of a given lattice and (ii) contributions due to non-zero mass $m_0$ of the $\eta^{\prime}$ mesons. This supports examining the nature of fluctuations of an ensemble of slab sub-volumes of a simulated lattice  in a time slice $(1/m_0, T-1/m_0)$ with or without the presence of statistical and systematic noises. Hereby, we focus on an ensemble of non-trivial local correlators with the topological charge density as the model embedding. Namely, under fluctuations of the slab dimensions, our consideration offers an intrinsic geometric classification of the stability, correlations and phases of an ensemble of slab sub-volumes of a simulated lattice $\mathbb{L}$. 

In general, for a given slab sub-volumes of an arbitrary simulated lattice, the end points $\{ t_i, t_{i+1} \}$ of its $i^{th}$ slice vary independently in the limit of small lattice spacing. Physically, given an ensemble of slab sub-volumes of an arbitrary simulated lattice  $\mathbb{L}$, it follows that our consideration allows a slab to dilate, viz. to contract and expand in time. To be precise, this enables one in determining the stability of an arbitrary QCD configuration as an ensemble of slab sub-volumes that are defined via the time slices $\{ t_i, t_j \}$ of $\mathbb{L}$ with the next label $j= i+1$. Hereby, without loss of the generality, we mar consider the fluctuation theory analysis of system with the topological susceptibility as the model embedding in the space of slab dimensions $\{ t_1, t_2 \}$ as per the framework of the slab sub-volume method, see \cite{Aoki} for details.
	
In the sequel, given an ensemble of slab-sub-volumes of an arbitrary simulated lattice  $\mathbb{L}$, we analyze the nature of fluctuations of the topological susceptibility through the slab sub-volume method  \cite{Aoki} as outlined in the previous section, viz. see Eqn.(\ref{slab}). Namely, for a given slab, let $q(t_i)=< Q^2_{slab} (t_i)>$, where $i= 1, 2$, the topological susceptibility reads as

\begin{equation}  \label{qstf}
\chi (t_1, t_2):= \frac{T}{V} \bigg( \frac{q(t_1) - q(t_2)}{t_1-t_2} \bigg)
\end{equation}
 
Hereby, as a special case, it follows that the topological susceptibility $\chi_t(t_1, t_2)$ reduces as a function of the difference $t_1 - t_2$ in the limit of an arbitrary large lattice. Namely, the topological susceptibility arises as a function of the difference $t_1 - t_2$ in the limit a fixed slab approximation of the given lattice. Indeed, such a particular class of lattices whose slab sub-volume dimensions $ t_1$ and $t_2$ have a constant difference, viz. the respective topological susceptibility depends on the parameter $t:= t_1 - t_2$ arise as a limiting case of our consideration. Geometrically, this can be viewed as a submersion from the space of the lattice parameters $\{t_1, t_2 \}$ to the set of real numbers. Algebraically, such a specification corresponds to a curve where one of the parameters is defined in terms of the other, that is, we have either the curve $t_2 \equiv t_2(t_1)$ or $t_1 \equiv t_1(t_2)$. The associated discussion of the results concerning the limiting infinitesimal fluctuations is relegated to Section 5 and that the numerical predictions of the model to Section 6.

Considering the fact that there can exist errors in observations of a long simulated lattice, that is, the concerned slabs can have a pair of variable dimensions $\{t_1, t_2\}$ that define an embedding as in Eqn.(\ref{manifold}), the flow components $\{ \chi_{t_i}:= \frac{\partial \chi}{\partial t_i}\ | \ i= 1, 2 \}$ reduce as 
\begin{eqnarray} \label{fc}
\chi_{t_1} &=& \frac{T}{V} \bigg( \frac{(t_1-t_2) q_{t_1}(t_1) - q(t_1)+ q(t_2)}{(t_1-t_2)^2} \bigg), \nonumber \newline \\
\chi_{t_2} &=& -\frac{T}{V} \bigg( \frac{(t_1-t_2) q_{t_2}(t_2) - q(t_1)- q(t_2)}{(t_1-t_2)^2} \bigg)
\end{eqnarray}

Hence, the flow equations $\{ \chi_{t_i}=0\ | \ i= 1, 2 \}$ yield the following pair of simultaneous partial differential equations

\begin{eqnarray} \label{fe}
q_{t_1}(t_1) &=& \frac{1}{t_1-t_2} \big( q(t_1)- q(t_2) \big),  \nonumber \newline \\
q_{t_2}(t_1) &=& \frac{1}{t_1-t_2} \big( q(t_1)- q(t_2) \big)
\end{eqnarray}

This give a pair of simultaneous ordinary differential equations when one of the variables $\{t_1, t_2 \}$ is held constant. For $t_1= \tilde{c}_1$ and $q(\tilde{c}_1)= \tilde{c}_2$, we have the following two point correlation function

\begin{eqnarray} \label{ql2}
q(t_2)= \tilde{c}_3( \tilde{c}_1- t)+ \tilde{c}_2,
\end{eqnarray}

where $\tilde{c}_3$ is an arbitrary integration constant. 
Similarly, for $t_2= \tilde{c}_4$ and $q(\tilde{c}_4)= \tilde{c}_5$, it follows that the two point correlation function reads as per the following expressions

\begin{eqnarray} \label{ql1}
q(t_1)= \tilde{c}_6 (t- \tilde{c}_4)- \tilde{c}_5,
\end{eqnarray}

where $\tilde{c}_6$ is another integration constant. 
Therefore, about the equilibrium defined by the flow equations $\chi_{t_1}= 0= \chi_{t_2}$, the two point correlation function $q(t)$ behaves as a straight line when it is viewed as a function of the slab dimension $t$. Namely, for $t_1= \tilde{c}_1$, from Eqn.(\ref{ql2}) we see that the value of $q(t_2)$ modulates linearly in $t_1$ with an arbitrary slop $- \tilde{c}_3$ and the intercept $\tilde{c}_1 \tilde{c}_3 + \tilde{c}_2$, and that for the $q(t_1)$, we see from Eqn.(\ref{ql1}) that it has the slop $\tilde{c}_6$ and the intercept $-\tilde{c}_5- \tilde{c}_4 \tilde{c}_6$ for $t_2= \tilde{c}_4$. In particular, as we approach the limiting value $t_1= \tilde{c}_1= t_2$ of the slab dimension on a simulated lattice, we notice from Eqn.(\ref{ql2}) that the two point correlation function $q(t_2)$ takes a constant value $\tilde{c}_2$. Similarly, in the limit of the slab dimension $t_1= \tilde{c}_4= t_2$,  it follows from Eqn.(\ref{ql1}) that we have $q(t_1)= -\tilde{c}_5$.  

In order to study the behavior of fluctuations about the flow equations as depicted in Eqn.(\ref{fe}), that is, the critical point of $\chi$ where the two point correlation function $q(t)$ has a linear behavior with respect to the slab dimension $t$, we need to compute the respective thickness capacities $\{ \chi_{t_i t_i}\ | \ i= 1, 2 \}$ and local correlation $\chi_{t_1 t_2}$ over the slab dimensions $\{t_1, t_2\}$. Given the flow component $\chi_{t_1}$ as in Eqn.(\ref{fc}), with $t= t_1-t_2$, we see that the $t_1$-fluctuation capacity $\chi_{t_1 t_1}$ can be expressed as

\begin{equation}\label{fc1}
\chi_{t_1 t_1}(t)= \frac{T}{V t^3} \big(a_1 t^2 - 2b_1t + 2d \big),
\end{equation}
where the coefficients $a_1, b_1$ and $d$ are given by
\begin{equation} \label{cfc}
a_1 = q_{t_1 t_1}(t_1),\ \ b_1= q_{t_1}(t_1),\ \ d= q(t_1)- q(t_2)
\end{equation}

Similarly, from the flow component $\chi_{t_2}$ as in 
Eqn.(\ref{fc}), we find that the $t_2$-fluctuation capacity $\chi_{t_2 t_2}$ reads as

\begin{equation}\label{fc2}
\chi_{t_2 t_2}(t)= \frac{T}{V t^3} \big(-a_2 t^2 - 2b_2t + 2d \big),
\end{equation}

where the coefficient $d$ and the variable $t$ remain the same as in the Eqn.(\ref{cfc}). The remaining coefficients $\{a_2, b_2 \}$ are as below

\begin{equation}
a_2 = q_{t_2 t_2}(t_2),\ \ b_2= q_{t_2}(t_2)
\end{equation}

Differentiating one of the flow components as in Eqn.(\ref{fc}), say the flow component $\chi_{t_i}$ with respect to $t_j$ for $i \ne j$, we find the following local statistical correlation

\begin{equation} \label{lcor}
 \chi_{t_1 t_2}(t)= \frac{T}{V t^3} \big(b_3t - 2d \big),
\end{equation}

where the coefficient $d$ and the variable $t$ remain the same as in the Eqn.(\ref{cfc}). The remaining coefficient $ b_3$ is given as

\begin{equation}
b_3= q_{t_1}(t_1)+ q_{t_2}(t_2)
\end{equation}

Under variations of the thicknesses $\{ t_1, t_2 \}$ over the surface $\mathcal{M}_2$, the associated fluctuation matrix $H$ arising by considering the function $\chi(t_1, t_2)$ as the embedding function is defined by

\begin{equation} \label{fm}
H:= \begin{bmatrix} \chi_{t_1t_1} & \chi_{t_1t_2} \\ \chi_{t_1t_2} & \chi_{t_2t_2} \end{bmatrix}
\end{equation}

With the above definition of the fluctuation matrix $H$ as in Eqn.(\ref{fm}), from Eqns.(\ref{fc1}, \ref{fc2}, \ref{lcor}), for all $(t_1, t_2) \in \mathcal{M}_2$, it follows that the corresponding determinant

\begin{equation} \label{det0}
\Delta(t_1, t_2):= \chi_{t_1t_1} \chi_{t_2t_2}  -\chi_{t_1t_2}^2
\end{equation}

can be expressed as per the following quadratic numerator formula

\begin{equation} \label{det}
\Delta(t)= \frac{T^2}{V^2 t^6} \big(\alpha_2 t^2 + 2 \alpha_1 t + 4 \alpha_0 \big),
\end{equation}

where the coefficient $\{ \alpha_i \ i = 0, 1, 2 \}$ are given by

\begin{eqnarray}
\alpha_0 &=& \big(q(t_1) - q(t_2)\big) \big(1- q(t_1) + q(t_2) \big),\nonumber \newline \\ 
\alpha_1 &=& 2 \big(q_{t_1}(t_1) + q_{t_2}(t_2) \big) \big(q(t_1) - q(t_2) \big)- q_{t_1}(t_1) - q_{t_2}(t_2)),\nonumber \newline \\
\alpha_2 &=& q_{t_1t_1}(t_1) - q_{t_2t_2}(t_2) - \big(q_{t_1}(t_1) + q_{t_2}(t_2) \big)^2
\end{eqnarray}

Under fluctuations of the slab dimensions $\{t_1, t_2\}$, the computation of the global correlation length follows by evaluating the determinant 

\begin{equation} \label{num}
N:= \begin{vmatrix}
\chi_{t_1t_1}  & \chi_{t_1t_2}  & \chi_{t_2t_2}  \\ 
\chi_{t_1t_1t_1}  & \chi_{t_1t_1t_2}  & \chi_{t_1t_2t_2}  \\ 
\chi_{t_1t_1t_2}  & \chi_{t_1t_2t_2}  & \chi_{t_1t_2t_2} 
\end{vmatrix}
\end{equation}

It is not difficult to see that the third derivatives of the topological susceptibility read as

\begin{eqnarray} \label{chiijk}
\chi_{t_1t_1t_1} &=& \frac{T}{V t^4} \big(c_1 t^3 + a_3t^2 - 6 d \big), \nonumber \newline \\ 
\chi_{t_1t_1t_2} &=& \frac{T}{V t^4} \big(a_3 t^2 - 6 b_2 t + 6d + 2(q_1-q_2) \big), \nonumber \newline \\ 
\chi_{t_1t_2t_2} &=& \frac{T}{V t^4} \big(b_3 t - 6d \big), \nonumber \newline \\ 
\chi_{t_2t_2t_2} &=& \frac{T}{V t^4} \big(- c_2 t^3 -3 a_5t^2 - 6 b_4 t+ 6 d \big)
\end{eqnarray}

Here, as a function of $t$, the coefficients $\{ c_1, c_2 \}$ of the qubic terms in $\chi_{t_it_jt_k}$, $\{ b_2, b_3, b_4 \}$ that of the quadratic terms, and $\{ a_3, a_4 \}$ that of the linear terms are given by

\begin{eqnarray}
c_1 &=& q_{111}, \nonumber \newline \\ 
c_2 &=& q_{222}, \nonumber \newline \\ 
b_2 &=& q_1, \nonumber \newline \\ 
b_3 &=& q_{22} + 5 q_{2} + 3 q_1, \nonumber \newline \\ 
b_4 &=& q_2, \nonumber \newline \\ 
a_3 &=& q_{11}, \nonumber \newline \\ 
a_4 &=& q_{22}
\end{eqnarray}

with the difference $d= q(t_1) - q(t_2)$.
Substituting the above expressions of $\chi_{t_it_j}$ as in Eqn.(\ref{fc1}, \ref{fc2}, \ref{lcor}) and that of the $\chi_{t_it_jt_k}$ as in Eqn.(\ref{chiijk}), the terms $\{N_i \ | \ i= 1, 2, 3 \}$ of the numerator 

\begin{equation}\label{num}
N=\chi_{11}N_1+ \chi_{12}N_2+ \chi_{22}N_3
\end{equation} 

of the Ruppeiner's scalar curvature $R$ can be expressed as the sum

\begin{eqnarray}\label{numc}
N_i= \frac{T^2}{V^2 t^8} \sum_{j=0}^{6} N_{ij}t^j,\ i= 1,2,3
\end{eqnarray}


It is not difficult to see that the coefficients $ N_{1j}$ simplify as
\begin{eqnarray}
N_{10} &=&   6d(\tilde{d}- 6d), \nonumber \newline \\
N_{11} &=&   6(2b_3 d- 6b_2 b_4- b_4 \tilde{d}), \nonumber \newline \\
N_{12} &=&   36 b_2 b_4+ 6 a_3 d -b_3^2- 3 a_4 \tilde{d}, \nonumber \newline \\
N_{13} &=&   18 a_4 b_2- -6 a_3 b_4- c_2 \tilde{d}, \nonumber \newline \\
N_{14} &=&   3(2 b_2 c_2- a_3 a_4), \nonumber \newline \\
N_{15} &=&   -a_3 c_2, \nonumber \newline \\
N_{16} &=& 0,
\end{eqnarray}

where $ \tilde{d}= 6d+ 2(q_1-q_2) $. Similarly, we observe that the coefficients $ N_{2j}$ are given by
\begin{eqnarray}
N_{20} &=&   6d(6d- \tilde{d}), \nonumber \newline \\
N_{21} &=&   36 (b_2- b_4) d - b_3 \tilde{d}, \nonumber \newline \\
N_{22} &=&   -6(2a_3+ 3a_4)d- 6b_2 b_4, \nonumber \newline \\
N_{23} &=&   a_3 (b_3+ 6b_4)- 6(c_1+ c_2)d, \nonumber \newline \\
N_{24} &=&   3(2b_4 c_1+ a_3 a_4), \nonumber \newline \\
N_{25} &=&   3a_4 c_1+ a_3 c_2, \nonumber \newline \\
N_{26} &=&   c_1 c_2
\end{eqnarray}

Furthermore, we find that the coefficients $ N_{3j}$ read as
\begin{eqnarray}
N_{30} &=&   36 d^2 - \tilde{d}^2, \nonumber \newline \\
N_{31} &=&   6(2b_2 \tilde{d}- b_3 d), \nonumber \newline \\
N_{32} &=&   -(6a_3 d+ 36 b_2^2+ 2 a_3 \tilde{d}), \nonumber \newline \\
N_{33} &=&   a_3(b_3 +12 b_2)-6c_1 d, \nonumber \newline \\
N_{34} &=&   b_3 c_1- a_3^2, \nonumber \newline \\
N_{35} &=&  0, \nonumber \newline \\
N_{36} &=& 0
\end{eqnarray}

Thus, it follows that the numerator $N$ of the scalar curvature reads as

\begin{equation}\label{num}
N= \frac{T^2}{V^2 t^{11}} \sum_{i=0}^{7} n_{i}t^i,
\end{equation} 

where the coefficients $\{ n_i \ i = 0, 1, \ldots, 7 \}$ simplify as per the following data

\begin{eqnarray} \label{nunc}
n_{0} &=&   2d(N_{10}- N_{20}+ N_{30}), \nonumber \newline \\
n_{1} &=&   b_3 N_{20}- 2(b_1 N_{10}+ b_2 N_{30})+ 2d (N_{11}- N_{21}+ N_{31}), \nonumber \newline \\
n_{2} &=&   a_1 N_{10}- a_2N_{30}+ b_3 N_{21}- 2(b_1 N_{11}+ b_2 N_{31})+ 2d (N_{12}- N_{22}+ N_{32}), \nonumber \newline \\
n_{3} &=&   a_1 N_{11}- a_2N_{31}+ b_3 N_{22}- 2(b_1 N_{12}+ b_2 N_{32})+ 2d (N_{13}- N_{23}+ N_{33}), \nonumber \newline \\
n_{4} &=&  a_1 N_{12}- a_2N_{32}+ b_3 N_{23}- 2(b_1 N_{13}+ b_2 N_{34})+ 2d (N_{14}- N_{24}+ N_{34}), \nonumber \newline \\
n_{5} &=&  a_1 N_{13}- 2 b_1 N_{14}- a_2 N_{33}-b_2N_{34}+ b_3 N_{24} + 2d (N_{15}-N_{25}), \nonumber \newline \\
n_{6} &=&  a_1 N_{14}- 2b_1 N_{15}- a_2 N_{34}+ b_3 N_{24}- 2d N_{26}, \nonumber \newline \\
n_{7} &=&  a_1 N_{15}+ b_3 N_{26}
\end{eqnarray}

Following the definition of the correlation area of a statistical system as the Ruppeiner scalar curvature of the fluctuation surface \cite{rupp1, tcb, Jencova, ourpaper, bntfr, bookbtg, bntngsb, book2018, fisica, aman, halfbpspaper, bntef, Weinhold1, Weinhold2, rupp2, Ruppeiner3, Ruppeiner4, McMillan, Simeoni, widom}  with its coordinates as the slab dimensions, from the Eqn.(\ref{cur}), it follows that above lattice QCD configuration with its embedding map as in Eqn.(\ref{slab}) possesses the following scalar curvature

\begin{equation} \label{curre}
R(t_1, t_2)= \frac{k_B}{2} \frac{N(t_1, t_2)}{\Delta(t_1, t_2)^2},  
\end{equation}

With the numerator $N$ as in Eqn.(\ref{num}), it follows that we have the following scalar curvature

\begin{equation} \label{curt}
R(t)= \frac{k_B}{2} \frac{V^2}{T^2}t \frac{\sum_{i=0}^{7} n_i t^i}{(\sum_{i=0}^{2} \alpha_i t^i)^2}
\end{equation}

where the coefficients $\{ n_i \ i = 0, 1, \ldots, 7 \}$ of the scalar curvature $R(t)$ remain the same as in Eqn.(\ref{nunc}) and $k_B$ is the standard Boltzmann's constant. 

Therefore, it follows that the system is globally non-interacting if we have a vanishing scalar curvature $R(t)=0$, that is, from Eqn.(\ref{curt}), the extent $t$ of the slab satisfies either has a vanishing value $t=0$, or it satisfies the following degree eight equation

\begin{equation}
\sum_{i=0}^{8} n_i t^i = 0
\end{equation}

Such a configuration corresponds to the pure designing of a slab concerning the technical instruments and measurements.
Herewith, from vanishing of the scalar curvature, we see that there are in total nine values of slab dimension $t \in \{0,\ t_i\ |\ i=1,2,\ldots,8 \} $ where the system becomes non-interacting without any global correlations.

On the other hand, the system under goes phase transitions when the scalar curvature $R(t)$ diverges, that is, the difference $t$ satisfies the following quadratic equation

\begin{equation}
\sum_{i=0}^{2} \alpha_i t^i = 0
\end{equation}

Hereby, given the denominator of $R$ as the square of $\Delta$ as in Eqn.(\ref{det}), we find that the phase transitions occur at the roots of the quadratic polynomial $\alpha_2 t^2 + 2 \alpha_1 t + 4 \alpha_0$. That is, the system becomes infinitely correlated when the slab dimensions $\{t_1, t_2\}$ are separated by

\begin{equation}
t_{\pm}= \frac{1}{2 \alpha_2}\bigg( -2 \alpha_1 \pm \sqrt{4 \alpha_1^2 - 16 \alpha_2 \alpha_0} \bigg)
\end{equation}  

This implies two values of $t=t_{\pm}$. Thus, in total there are two values of slab dimension $t \in \{t_+, t_- \} $ where there exist possibilities of a large global correlation and phase transitions. In the sequel, in order to make its comparison with the ChPT formulation of the topological susceptibility fluctuations, associated physical phenomena and validation of the model, we specialize to an ensemble of infinitesimal slab sub-volumes of an arbitrary simulated lattice.

\section{Infinitesimal Slab Fluctuations}

In this section, we discuss fluctuation theory analysis of the two point correlation function $q$ as in Eqn.(\ref{qstf}) over an infinitesimal slab when its dimensions $\{t_1, t_2\}$ are separated by an infinitesimal interval $h:= t_1-t_2$.

Following the motivations from the previous section, recall that the topological susceptibility based on the slab sub-volumes of an arbitrary simulated lattice as in Eqn.(\ref{slab}) can be expressed as

\begin{equation}
\chi(t_1, t_2)= C \frac{q(t_1) - q(t_2)}{t_1-t_2},
\end{equation}  

where the prefactor $C$ of the the topological susceptibility $\chi$ is defined as

\begin{equation} \label{Cp}
C=T/V
\end{equation}  

In order to make comparison of the slab sub-volume predictions with that of the ChPT formulation, we may proceed as follows. Given a constant topological susceptibility $\chi=k$ over a lattice $\mathbb{L}:= \{t_i\ |\ i= 1, 2, 3, \ldots, N\}$ and the prefactor $C$ as in the above Eqn.(\ref{Cp}), we have the following iterative formula for the two point correlation 

\begin{equation} \label{2ptf}
q(t_2)= q(t_1) - (t_1-t_2) \frac{k}{C}
\end{equation}

as a function of the forgoing two point function $q(t_1)$ and the considered lattice points $\{t_1, t_2\}$. In the sequel, considering uncertainties in the slab dimensions $\{t_1, t_2\}$, we offer the fluctuation theory analysis of $q$ when the slab parameters $t_1$ and $t_2$ are close to each other. To do so, let us consider an infinitesimal slab with its dimension $h= t_1 - t_2 \rightarrow 0$, that is, we have $t_2= t_1 - h$. Indeed, we can consider any two neighboring points $(t_i, t_{i+1})$ on the simulated lattice $\mathbb{L}$ for which the subsequent analysis remains intact via the replacement of $1\rightarrow i$ and $2\rightarrow i+1$. In this case, it follows that we have 

\begin{equation}
q(t_1-h)= q(t_1) - (t_1-t_2) \frac{k}{C}
\end{equation}

Using Taylor series expansion, we see that the two point function $q(t_1)$ satisfies

\begin{equation}
-h q_{t_1}+ \frac{h^2}{2} q_{t_1 t_1}- \frac{h^3}{6} q_{t_1 t_1 t_1} \ldots = - \frac{k}{C}(t_1-t_2),
\end{equation}

where the subscript $t$ on $q$ denotes its partial derivative with $t$. Therefore, to find the critical points of $q(t_1)$, we may consider the flow equation $\frac{\partial q}{\partial t_1}=0$. In the sequel, we perform our analysis in the light of quadratic fluctuations. With $h= t_1-t_2$, this implies that $q(t_1)$ satisfies the following second order differential equation

\begin{equation}
\frac{\partial^2 q}{\partial t_1^2}= - \frac{2k}{C} \frac{1}{ (t_1-t_2) }
\end{equation}

Integrating both side with respect to $t_1$ twice and considering $t_2$ as a constant, we get the following two point correlation function

\begin{equation} \label{q1}
q(t_1)=  C_2+ C_1 t_1 - \frac{2k}{C} (t_1-t_2) \ln (\frac{ (t_1-t_2) }{e}),
\end{equation}

where $C_1$ and $C_2$ are constants of the integration. From Eqn.(\ref{2ptf}), we find that the subsequent two point correlation function $q(t_2)$ at $t=t_2$ reads as

\begin{equation}\label{q2}
q(t_2)=  C_2+ (C_1- \frac{k}{C}) t_1+ \frac{k}{C} t_2 - \frac{2k}{C} (t_1-t_2) \ln (\frac{ (t_1-t_2) }{e})
\end{equation}

Given the above expressions of the two point functions $\{q(t_1), q(t_2)\}$ as in Eqns.(\ref{q1}, \ref{q2}) as the functions of $\{t_1, t_2\}$, the flow components are given by

\begin{eqnarray}
\frac{\partial q}{\partial t_1}&=& C_1- \frac{2k}{C} \ln (t_1-t_2), \nonumber \newline \\  
\frac{\partial q}{\partial t_2}&=& \frac{k}{C}+ \frac{2k}{C} \ln (t_1-t_2)
\end{eqnarray}

Thus, the flow equation defined as $\frac{\partial q}{\partial t_1}= 0= \frac{\partial q}{\partial t_2}$ yields the critical values of the slab dimensions $\{t_1, t_2\}$ as per the following

\begin{equation} \label{icritical}
t_1- t_2 =  
\begin{cases} 
\exp (-\frac{1}{2}), & \text{if } \frac{k}{C} \ne 0 \\
\exp (\frac{C C_1}{2k}), & \text{otherwise }
\end{cases}   
\end{equation}

Hereby, we see that the critical values of $\{t_1, t_2\}$ satisfy positivity of the step size $h$, that is, from Eqn.(\ref{icritical}) we have a forward movement on the simulated lattice $\mathbb{L}$ with its step size

\begin{equation}
h:= t_1- t_2 >0,\ \forall \ (t_1, t_2) \in \mathcal{M}_2
\end{equation}

In order to check the stability of the two point correlation function $q$ as a function of the slab dimensions $\{t_1, t_2\} $, we need to evaluate its second derivatives under variations of $\{t_1, t_2\}$. In this case, it follows that we have an identical pair of fluctuation capacities

\begin{eqnarray} \label{ifcapa}
\frac{\partial^2 q}{\partial t_1^2}=  -\frac{2k}{C} \frac{1}{t_1-t_2}= 
\frac{\partial^2 q}{\partial t_2^2}
\end{eqnarray}

The local correlation function in the space of the slab dimensions $\{t_1, t_2\} $ is given by 

\begin{eqnarray} \label{ilcor}
\frac{\partial^2 q}{\partial t_1 \partial t_2}=  \frac{2k}{C} \frac{1}{t_1-t_2}
\end{eqnarray}

Thus, from Eqns.(\ref{ifcapa}, \ref{ilcor}) it follows with the definition of the fluctuation matrix 

\begin{equation} \label{ifm}
H_{inf}:= \begin{bmatrix} q_{11} & q_{12} \\ q_{12} & q_{22} \end{bmatrix},
\end{equation}

 that under variations of the infinitesimal slab dimensions $\{t_1, t_2\} $, the  limiting fluctuation determinant vanishes identically, that is,  we have

\begin{equation} \label{idet}
\Delta_{inf}:= q_{11} q_{22}-  q_{12}= 0,\ \forall \ (t_1, t_2) \in \mathcal{M}_2
\end{equation}

Herewith, we notice that the two point correlation function $q$ shows a degenerate behavior under variations of the $\{t_1, t_2\}$ as the determinant as depicted in Eqn.(\ref{idet}) take null value at the critical values of $(t_1, t_2)$ as in Eqn.(\ref{icritical}). Further, for all positive topological susceptibility $k>0$, it is worth mentioning that both the fluctuation capacities $\{q_{11}, q_{22}\}$ as in Eqn.(\ref{ifcapa}) take a negative value at the critical choice of the slab dimensions $\{t_1, t_2\} $. Namely, for the critical choices of $(t_1, t_2) \in \mathcal{M}_2$ as in Eqn.(\ref{icritical}), we see that  $\{q_{11}, q_{22}\}$ read as

\begin{eqnarray} 
\frac{\partial^2 q}{\partial t_1^2}=  -\frac{2k}{C} \exp(\tilde{C})= 
\frac{\partial^2 q}{\partial t_2^2},
\end{eqnarray}

where the exponent $\tilde{C}$ is defined as per the following identification

\begin{equation} 
\tilde{C} =  
\begin{cases} 
\frac{1}{2}, & \text{if } \frac{k}{C} \ne 0 \\
- \frac{C C_1}{2k}, & \text{otherwise }
\end{cases}   
\end{equation}

At the above critical values of $(t_1, t_2)$ as in Eqn.(\ref{icritical}), we observe that the fluctuation determination $\Delta_{inf}$ 
%
%
always vanishes. Therefore, for all $(t_1, t_2) \in \mathcal{M}_2$, we have a degenerate surface for the limiting statistical system corresponding to an infinitesimal slab under fluctuations of its dimensions $\{t_1, t_2\}$. At this juncture, in order to investigate the global correlation length $l_{inf}$, we need to evaluate the following determinant 

\begin{equation} \label{inmatrix}
N_{inf}:= \begin{vmatrix}
q_{11}  & q_{12}  & q_{22}  \\ 
q_{111}  & q_{112}  & q_{122}  \\ 
q_{112}  & q_{122}  & q_{222} 
\end{vmatrix}
\end{equation}

In this case, it is not difficult to see that the two point correlation function $q$ has the following third order derivatives

\begin{eqnarray} \label{ithirdderi}
q_{111} &=& \frac{2 k}{C} \frac{1}{(t_1-t_2)^2} = q_{122}, \nonumber \newline \\ q_{112}&=& -\frac{2 k}{C} \frac{1}{(t_1-t_2)^2}= q_{222}
\end{eqnarray}

From Eqns.(\ref{ifcapa}, \ref{ilcor}, \ref{ithirdderi}), it follows further that the determinant $N_{inf}$ in above in Eqn.(\ref{inmatrix}) vanishes identically since its first and third columns are identical in their values, that is, we have $N_{inf}=0$. Now, following the formula of the Ruppeiner scalar curvature invariant as in Eqn.(\ref{cur}), we find that the limiting infinitesimal slab fluctuations yield an indeterminate scalar curvature invariant, that is, under fluctuations over its dimensions $\{t_1, t_2\}$, we have that

\begin{equation}
R_{inf}(t_1, t_2):= \frac{k_B}{2} \frac{N_{inf}}{\Delta_{inf}^2}
\end{equation}

take a $0/0$ form for all $(t_1, t_2) \in \mathcal{M}_2$. Thus, the ensemble of infinitesimal slab sub-volumes of a simulated lattice $\mathbb{L}$ yields an ill-defined statistical basis. Physically, the fluctuations over the dimensions $t_1$ and $t_2$ of an infinitesimal slab correspond to a degenerate statistical system. This shows that the fluctuation theory analysis of the slab sub-volume method and its equivalence with the corresponding fluctuations of the ChPT formulation as performed in section $3$ requires a refined examination. Such investigations we leave open for a future research.

In the light of the foregoing section, it is worth mentioning that the slab sub-volume method is a generalized formulation for computing the topological susceptibility of $2+1$ flavor QCD on a simulated lattice. In this setup, we find that the ensemble of finite slab sub-volumes of an arbitrary lattice have a well controlled behavior under fluctuations of the slab dimensions $\{t_1, t_2\}$. 
In reference to the statistical uncertainties, topology freezing and systematic effects, our analysis opens new avenues towards the examination of an ensemble of infinitesimal slab sub-volumes of an arbitrary simulated lattice, namely, to examine the question whether the slab method yields identical intrinsic geometric structures to its ChPT counterparts. It would be interesting to explore the limit of $t_1$ and $t_2$ that corresponds to ChPT ensemble fluctuations in the space of the quark mass and the mass scale pertaining to renormalized pion masses.

\section{Discussion of the Results}

In this section, we provide physical implications of the results concerning the ChPT formulation and slab sub-volume fluctuations while computing the topological susceptibility fluctuations. Firstly, we give a qualitative discussion of results when the model parameters are allowed to fluctuate in the light of ChPT formulation. 
In the sequel, we describe physical implications of the slab sub-volume fluctuations of computing the topological susceptibility and compare them with the existing models such as the aforementioned ChPT predictions, and that of the the ALPHA and JLQCD collaborations.

\subsection{Numerical Predictions}

In this subsection, we discuss qualitative implication of the results concerning the topological susceptibility fluctuations as a function of the model parameters as the (i) quark mass and the mass scale that corresponds to renormalized pion masses for the ChPT formulation and (ii) slab dimensions for the slab sub-volume method. Subsequently, we compare the nature of fluctuations and associated ensemble stability properties by considering the respective topological susceptibility as the model embedding function.

\subsubsection{ChPT Predictions}

Following the observations as in section $3$, we offer the qualitative behavior of the topological susceptibility and associated fluctuation quantities as a function of the quark mass $m$ and the mass scale $M$ that corresponds to renormalized pion masses. To do so, lets fix the parameters of the ChPT model as follows. With the choice of parameters in the $\overline{MS}$ scheme \cite{Aoki} as

\begin{equation}
\Sigma^{\overline{MS}}(2\ GeV)= (270\ MeV)^3,\ F_{\pi} = 130\ MeV,\ l = -0.021,
\end{equation}

the model parameters $\{a, b, c, d\}$ as in Eqn.(\ref{notations}) respectively read as
 
 \begin{equation} \label{rcv}
 a= 9841500\ MeV^3,\ b= 12898\ MeV^2,\ c= 2329\ MeV,\ d= -56972\ MeV^2 
\end{equation}
 
 \begin{center}
	\begin{figure}
		\hspace*{1.0cm} \vspace*{-7.6cm}
		\includegraphics[width=13.0cm,angle=0]{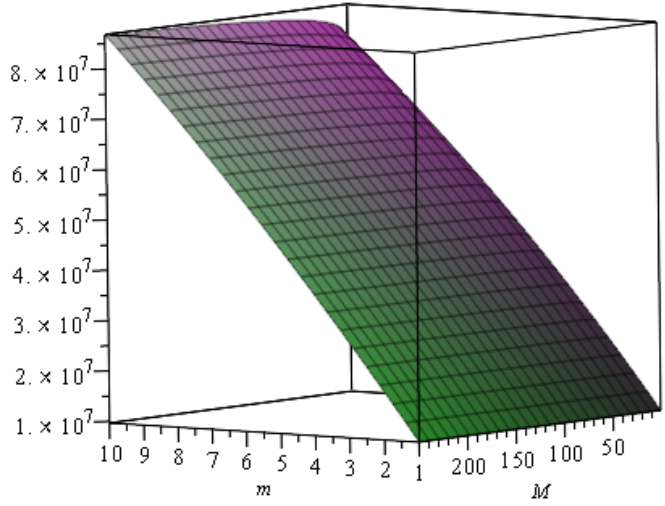}
		\caption{The topological susceptibility $\chi(m, M)$ plotted as a function of the quark mass $m$ on $X$-axis and the mass scale $M$ that corresponds to renormalized pion masses on $Y$-axis, describing the nature of global topological interactions along the $Z$-axis by considering $\chi(m, M)$ as the embedding function of the fluctuation configuration under variations of $m$ and $M$.} \label{fig1}
		\vspace*{-0.01cm}
	\end{figure}
\end{center}

With the above values of the parameters as in Eqn.(\ref{rcv}), the ChPT formulation based topological susceptibility $\chi(m,M)$ is depicted in Fig.(\ref{fig1}) as a function of the quark mass $m$ on $X$ and the mass scale $M$ that corresponds to renormalized pion masses on $Y$-axis respectively. Hereby, we see that $\chi(m,M)$ has an amplitude of $8 \times 10^{7}$. This describes the nature of global topological interactions by considering $\chi(m,M)$ as the embedding function on the surface of fluctuations defined by variations of $m$ and $M$ in their respective ranges $(1, 10)$ and $(1, 250)$. In particular, we observe that $\chi(m,M)$ has an increasing behavior with respect to quark mass $m$ for a given values of the mass scale $M$ that corresponds to renormalized pion masses in $0<M<250$.  From, Fig.(\ref{fig1}), we see that $\chi(m,M)$ has a smooth flow without fluctuations.

\begin{center}
	\begin{figure}
		\hspace*{1.0cm}\vspace*{-7.2cm}
		\includegraphics[width=12.0cm,angle=0]{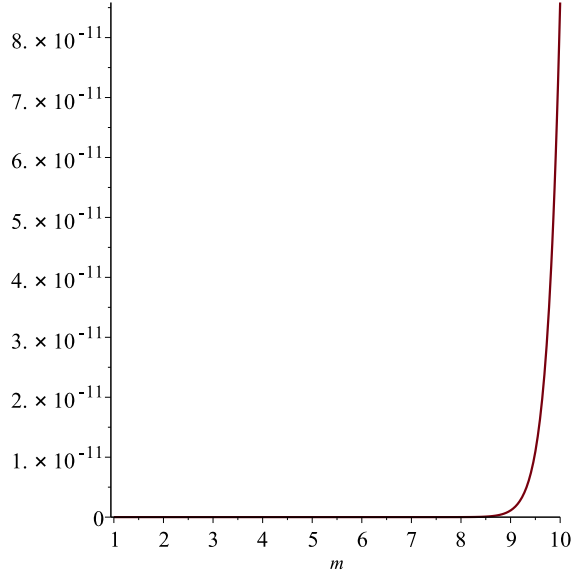}
		\caption{The mass scale $M$ that corresponds to renormalized pion masses as a function of the quark mass $m$ about the equilibrium defined by the flow equations plotted as a function of the quark mass $m$ on $X$-axis and the mass scale $M$ that corresponds to renormalized pion masses on $Y$-axis, describing the nature of global topological interactions by considering $\chi(m,M)$ as the embedding function of the fluctuation configuration under variations of the quark mass $m$ in $(0, 10)$.} \label{Mp10}
		\vspace*{-0.01cm}
	\end{figure}
\end{center}

\begin{center}
	\begin{figure}
		\hspace*{1.0cm}\vspace*{-7.2cm}
		\includegraphics[width=12.0cm,angle=0]{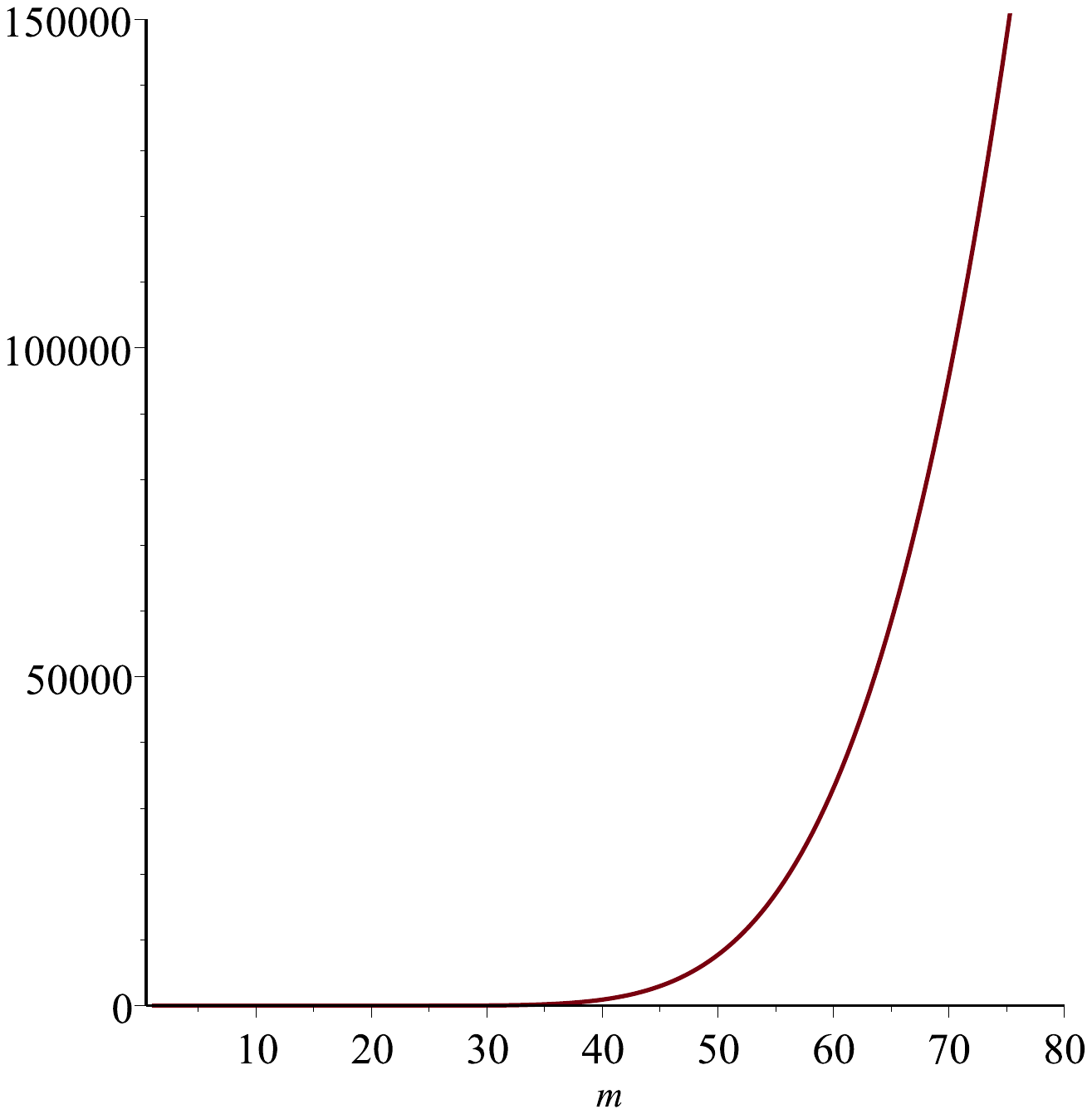}
		\caption{The mass scale $M$ that corresponds to renormalized pion masses as a function of the quark mass $m$ about the equilibrium defined by the flow equations plotted as a function of the quark mass $m$ on $X$-axis and the mass scale $M$ that corresponds to renormalized pion masses on $Y$-axis, describing the nature of global topological interactions by considering $\chi(m,M)$ as the embedding function of the fluctuation configuration under variations of the quark mass $m$ in $(0, 100)$.} \label{Mp100}
		\vspace*{-0.01cm}
	\end{figure}
\end{center}

Herewith,  we observe that the mass scale $M$ that corresponds to renormalized pion masses as a function of the quark mass $m$ about the equilibrium defined by the flow equations $\chi_m=0=\chi_M$ shows an increasing behavior, cf. in Fig.(\ref{Mp10}) and Fig.(\ref{Mp100}). Notice that $M$ has different amplitudes for different values of the quark mass. Fig.(\ref{Mp10}) shows the behavior of the mass scale $M$ that corresponds to renormalized pion masses by considering $\chi(m,M)$ as the embedding function of the fluctuation configuration when the quark mass $m$ varies in $(0, 10)$. In this case, we see that $M$ grows rapidly about $m=9$ and shoots up to an amplitude of the order $10^{-11}$ for the quark mass $9<m<10$.
Similarly, the behavior of the mass scale $M$ that corresponds to renormalized pion masses is shown in Fig.(\ref{Mp100}) when the quark mass $m$ varies in the range $(0, 100)$ where the topological susceptibility $\chi(m,M)$ is considered as the embedding function on the fluctuation manifold $\mathcal{M}_2$. 
From Fig.(\ref{Mp100}), we observe that $M$ starts growing rapidly about $m=40$, whereby it approaches to a large amplitude of $150000$ when $m \approx 80$.

\begin{center}
	\begin{figure}
		\hspace*{1.0cm}\vspace*{-7.6cm}
		\includegraphics[width=12.0cm,angle=0]{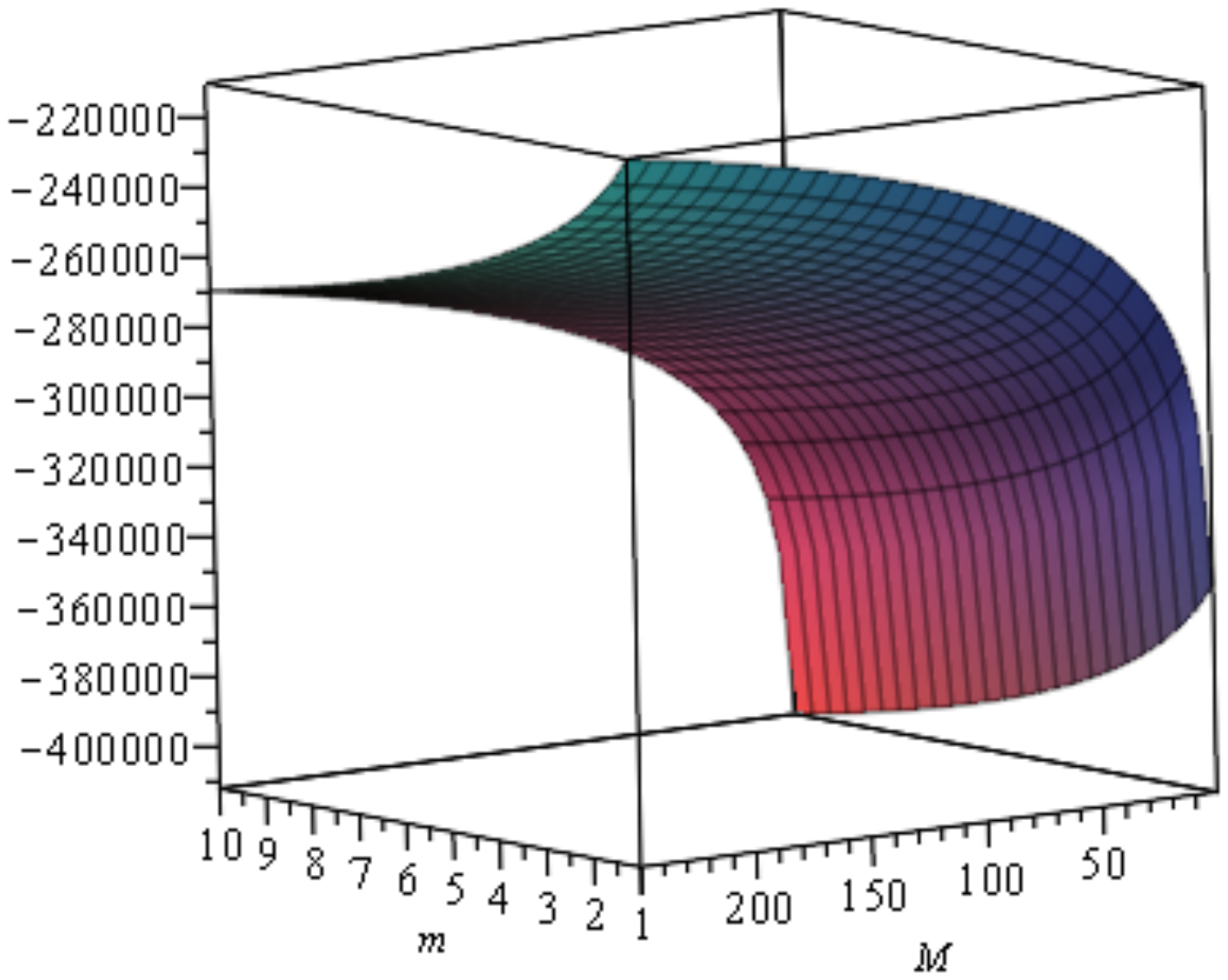}
		\caption{The quark mass capacity $\chi_{mm}(m,M)$ plotted as a function of the quark mass $m$ on $X$-axis and the mass scale $M$ that corresponds to renormalized pion masses} on $Y$-axis, describing the nature of local stability mediated via  topological interactions along the $Z$-axis by considering the topological susceptibility $\chi(m,M)$ as the embedding function on the fluctuation configuration under variations of $m$ and $M$. \label{fig2}
		\vspace*{-0.01cm}
	\end{figure}
\end{center}

By considering the topological susceptibility $\chi(m,M)$ as the  embedding function of the surface of fluctuations of the quark mass $m$ and the mass scale $M$ that corresponds to renormalized pion masses, the ChPT formulation based quark mass capacity $\chi_{mm}(m,M)$ is depicted in Fig.(\ref{fig2}). Herewith, we notice that $\chi_{mm}(m,M)$ has a negative amplitude of the order $220000$. This describes the nature of local topological susceptibility fluctuations  in the sapace of the quark mass $m$ and the mass scale $M$ corresponding to renormalized pion masses in their respective range $(1, 10)$ and $(1, 250)$. Hereby, we find that $\chi(m,M)$ has a decreasing behavior with respect to the mass scale $M$ that corresponds to renormalized pion masses for a given values of the quark mass $0<m<10$. 

\begin{center}
	\begin{figure}
		\hspace*{1.0cm}\vspace*{-7.6cm}
		\includegraphics[width=12.0cm,angle=0]{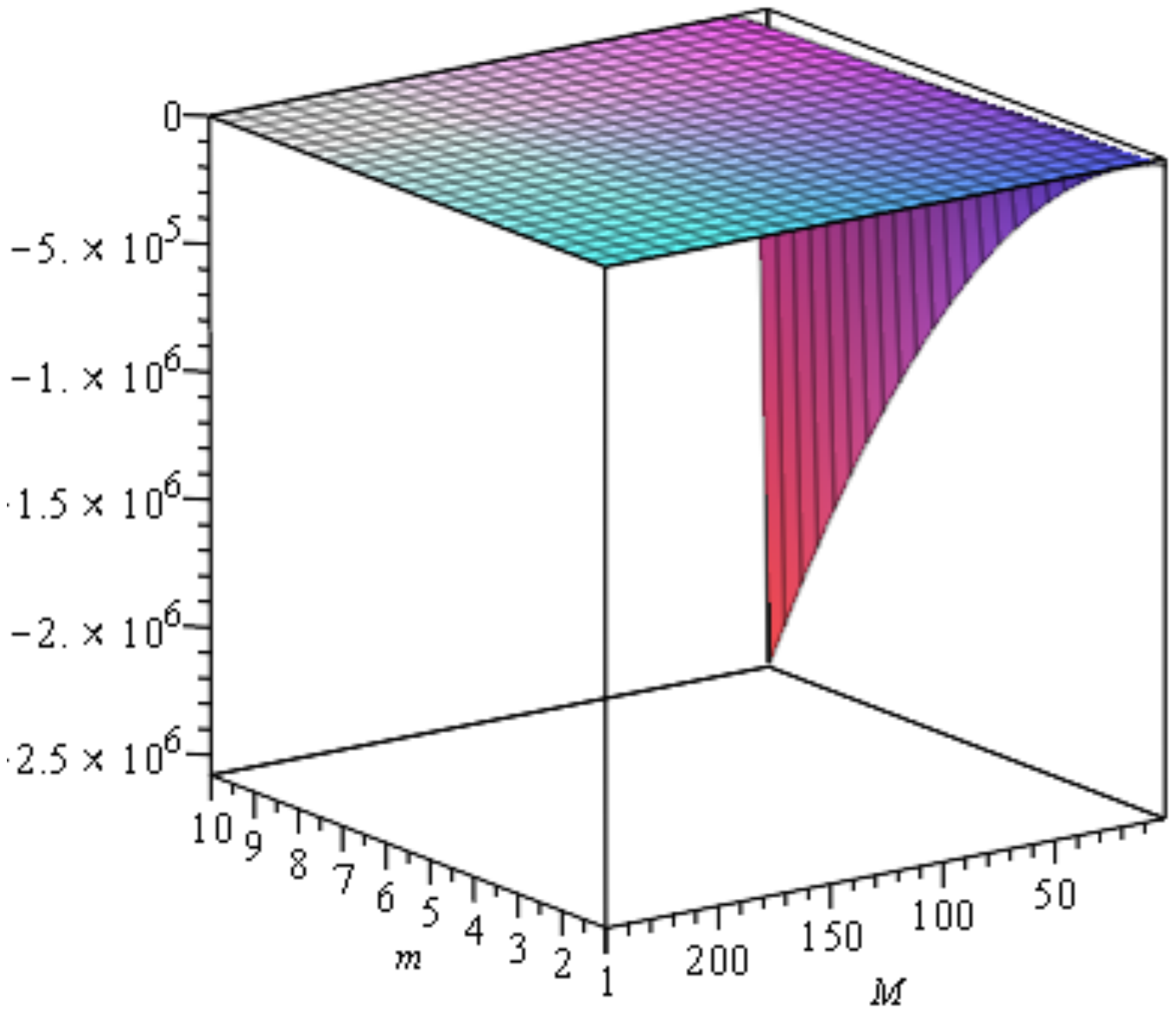}
		\caption{The pion mass capacity $\chi_{MM}(m,M)$ plotted as a function of the quark mass $m$ on $X$-axis and the mass scale $M$ that corresponds to renormalized pion masses on $Y$-axis describing the nature of local stability mediated via  topological interactions along the $Z$-axis by considering the topological susceptibility $\chi(m,M)$ as the embedding function on the fluctuation configuration under variations of $m$ and $M$.} \label{fig3}
		\vspace*{-0.01cm}
	\end{figure}
\end{center}

The physical mass scale capacity $\chi_{MM}(m,M)$ with the topological susceptibility $\chi(m,M)$ as the  embedding function of the surface of the quark mass $m$ and the mass scale $M$ that corresponds to renormalized pion masses is given in Fig.(\ref{fig3}). In this case, it follows that $\chi_{MM}(m,M)$ has a negative amplitude of its value $-2.5 \times 10^{-6}$. This describes the nature of local topological susceptibility fluctuations in the space of the quark and mass $m$ and the mass scale $M$ corresponding to renormalized pion masses in their respective range $(1, 10)$ and $(1, 250)$. Hereby, we observe that $\chi_{MM}(m,M)$ has a decreasing behavior with respect to a large mass scale $M$ for the quark mass $0<m<10$. We equally notice that there are fluctuations only for small values of the mass scale $M$ approaching the origin of the surface of fluctuations. 

\begin{center}
	\begin{figure}
		\hspace*{1.0cm}\vspace*{-7.6cm}
		\includegraphics[width=12.0cm,angle=0]{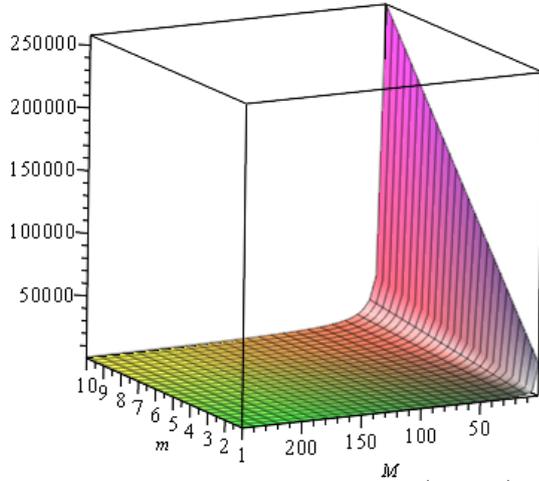}
		\caption{The parametric correlation $\chi_{mM}(m,M)$ plotted as a function of the quark mass $m$ on $X$-axis and the mass scale $M$ that corresponds to renormalized pion masses on $Y$-axis, describing the nature of local stability mediated via topological interactions along the $Z$-axis by considering the topological susceptibility $\chi(m,M)$ as the embedding function on the fluctuation configuration under variations of $m$ and $M$.} \label{fig4}
		\vspace*{-0.01cm}
	\end{figure}
\end{center}

The parametric correlation $\chi_{mM}(m,M)$ with the topological susceptibility $\chi(m,M)$ as the embedding function is plotted in Fig.(\ref{fig4}) as a function of the quark mass $m$ on $X$-axis and the mass scale $M$ that corresponds to renormalized pion masses on $Y$-axis. Considering the fluctuation surface with it coordinates as the quark mass $m$ and mass scale $M$ corresponding to renormalized pion masses, it follows that $\chi_{mM}(m,M)$ has a positive amplitude of the value $2.5 \times 10^{-6}$. This describes the nature of local statistical correlations in the space of the quark mass $m$ and the mass scale $M$  in their respective ranges $(1, 10)$ and $(1, 250)$. Herewith, we see that $\chi_{mM}(m,M)$ has an increasing behavior with respect to the quark mass $0<m<10$ for a small value of the mass scale $M$ that corresponds to renormalized pion masses. We equally see that there are fluctuations in the system for a large value of the mass scale $M$.

\begin{center}
	\begin{figure}
		\hspace*{1.0cm}\vspace*{-7.6cm}
		\includegraphics[width=12.0cm,angle=0]{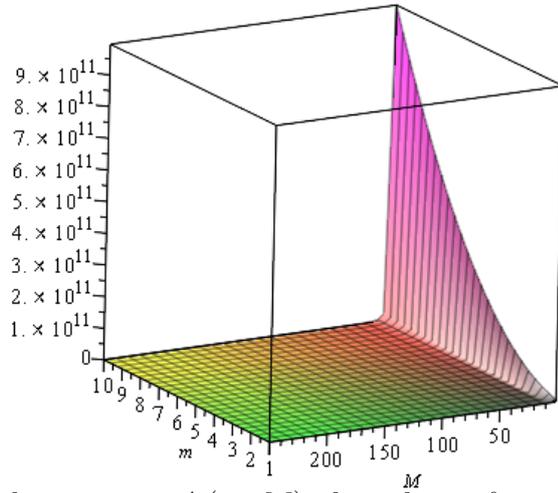}
		\caption{The determinant $\Delta(m,M)$ plotted as a function of the quark mass $m$ on $X$-axis and the mass scale $M$ that corresponds to renormalized pion masses on $Y$-axis, describing the nature of global stability mediated via the topological interactions along the $Z$-axis by considering the topological susceptibility $\chi(m,M)$ as the embedding function on the fluctuation configuration $\mathcal{M}_2$ under variations of the parameters $m$ and $M$.} \label{fig5}
		\vspace*{-0.01cm}
	\end{figure}
\end{center}

The fluctuation determinant $\Delta(m,M)$ is plotted in Fig.(\ref{fig5}) by considering the topological susceptibility $\chi(m,M)$ as the  embedding function on the fluctuation surface of the quark mass $m$ and  the mass scale $M$ that corresponds to renormalized pion masses. In this case, we see that  $\Delta(m,M)$ has the opposite behavior to that of the mass scale fluctuation capacity $\chi_{MM}(m,M)$. Namely, we find that  $\Delta(m,M)$ has a positive amplitude of its value $9 \times 10^{11}$. This describes the global nature of the topological susceptibility fluctuations in the space of the quark mass $m$ and  the mass scale $M$ corresponding to renormalized pion masses in their range $(1, 10)$ and $(1, 250)$ respectively. Hitherto, we notice that $\Delta(m,M)$ has an increasing behavior with respect to the quark mass $0<m<10$ for a small value of the mass scale $M$ that corresponds to renormalized pion masses. In addition, we  notice that there are nonlinear fluctuations in the limit of a large mass  scale $M\rightarrow 50$. It is worth mentioning that the amplitude of $\Delta(m,M)$ further grows as we move away from the origin of the fluctuation surface of the parameters $m$ and $M$. 

\subsubsection{Slab Sub-volume Predictions}

Following the observations as in Section $5$, in this subsection, we discuss the qualitative behavior of the slab sub-volume method based fluctuation quantities emerging from the two point correlation function as the model embedding function. In order to provide a graphical description of the slab sub-volume method based quantities, we need to fix its model parameters. This may be done as follows.  With the choice \cite{Aoki} of parameters 

\begin{equation} \label{spv}
T= 40,\ V= 32\ \mbox{and}\ 48, F_{\pi} = 30 MeV,\ M_{\pi} = 230 MeV,\ \chi:=k = 0.227 M_{\pi}^2 F_{\pi}^2,
\end{equation}
the model parameter $C$ as in Eqn.(\ref{Cp}) reads as $C= 5/4 \ \mbox{and}\ 5/12$ respectively fo $V= 32\ \mbox{and}\ 48$. What follows further, let we choose the arbitrary integration constants $\{C_1, C_2 \}$ as $C_1=1$ and $C_2=0$ in order to offer the qualitative behavior of fluctuation quantities. 

\begin{center}
	\begin{figure}
		\hspace*{1.0cm}\vspace*{-7.6cm}
		\includegraphics[width=13.0cm,angle=0]{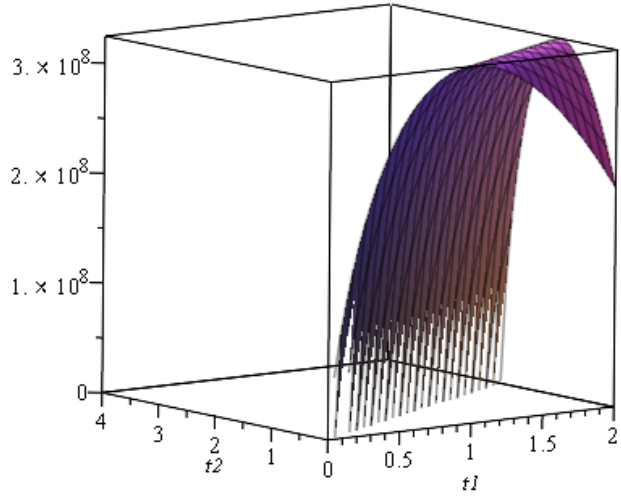}
		\caption{The two point correlation function $q_1(t_1,t_2)$ plotted as a function of the slab dimensions $\{ t_1, t_2 \}$ on $X$-axis and $Y$-axis respectively and the nature of the limiting $q_1$ along the $Z$-axis by considering it as the embedding function of the parametric figuration under fluctuations of the slab dimensions $t_1$ and $t_2$ with the lattice volume $V=32$.} \label{fig6}
		\vspace*{-0.01cm}
	\end{figure}
\end{center}

With the above values of the slab parameters as in Eqn.(\ref{spv}), the slab sub-volume formulation based two point correlation function $q_1(t_1,t_2)$ for a given value of the topological susceptibility $\chi:=k$ is depicted in Fig.(\ref{fig6}) as a function of the slab dimensions $t_1$ on $X$-axis and $t_2$ on $Y$-axis respectively for the lattice volume $V=32$. Hereby, we see that the amplitude of $q_1(t_1,t_2)$ attains the maximum value of $2 \times 10^{8}$ and the minimum value of $-1 \times 10^{8}$. This describes the nature of the two point correlation function $q_1(t_1,t_2)$ interactions by considering it as the embedding function on the surface of fluctuations under variations of the slab dimensions $t_1$ and $t_2$ in their respective ranges $(0, 2)$ and $(0, 4)$, also see \cite{Aoki} for an associated estimation. 

\begin{center}
	\begin{figure}
		\hspace*{1.0cm}\vspace*{-7.6cm}
		\includegraphics[width=13.0cm,angle=0]{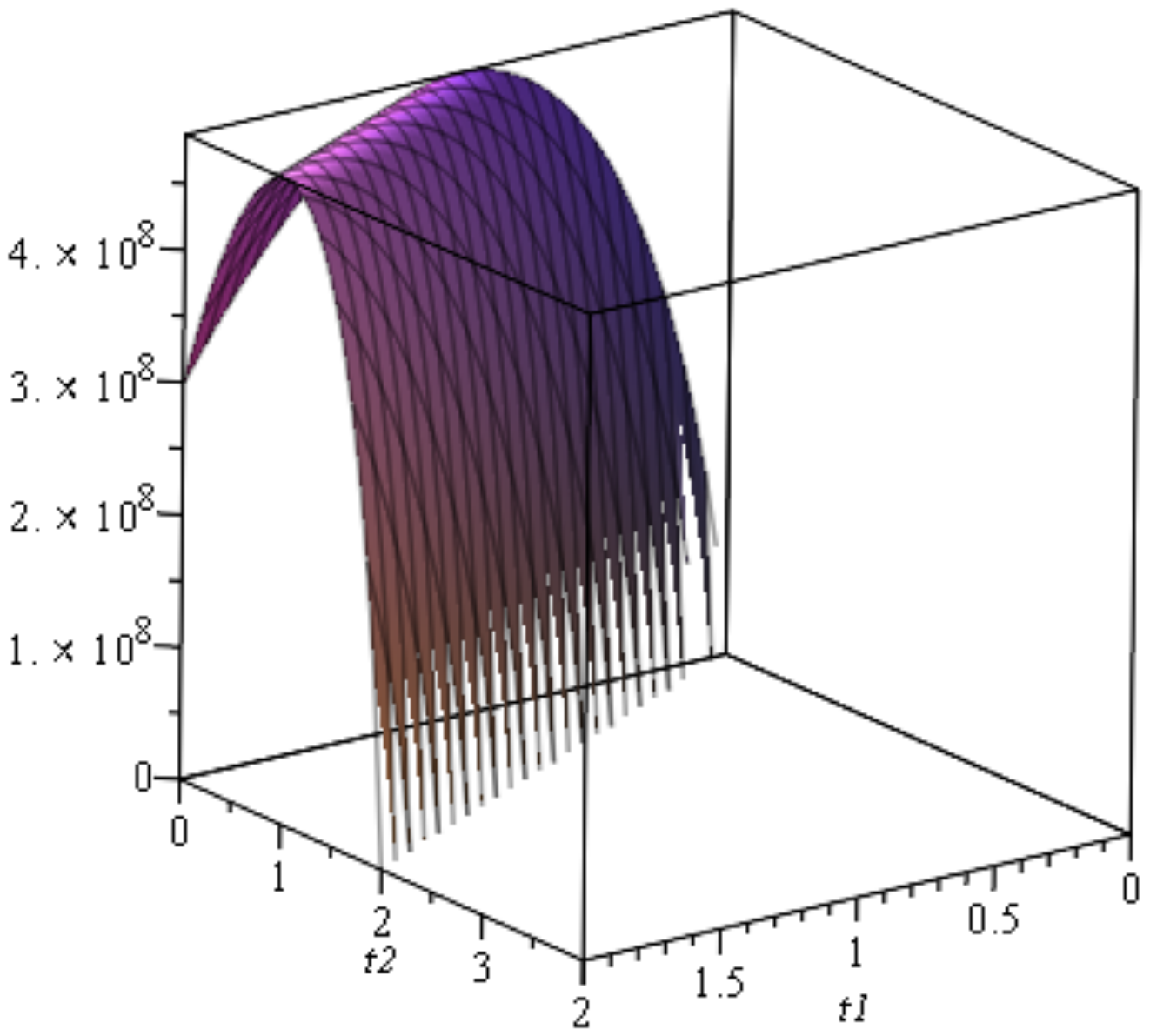}
		\caption{The two point correlation function $q_1(t_1,t_2)$ plotted as a function of the slab dimensions $\{ t_1, t_2 \}$ on $X$-axis and $Y$-axis respectively and the nature of the limiting $q_1$ along the $Z$-axis by considering it as the embedding function of the fluctuation configuration under variations of $t_1$ and $t_2$ with the lattice volume $V=48$.} \label{fig6a}
		\vspace*{-0.01cm}
	\end{figure}
\end{center}

\begin{center}
	\begin{figure}
		\hspace*{1.0cm}\vspace*{-7.6cm}
		\includegraphics[width=13.0cm,angle=0]{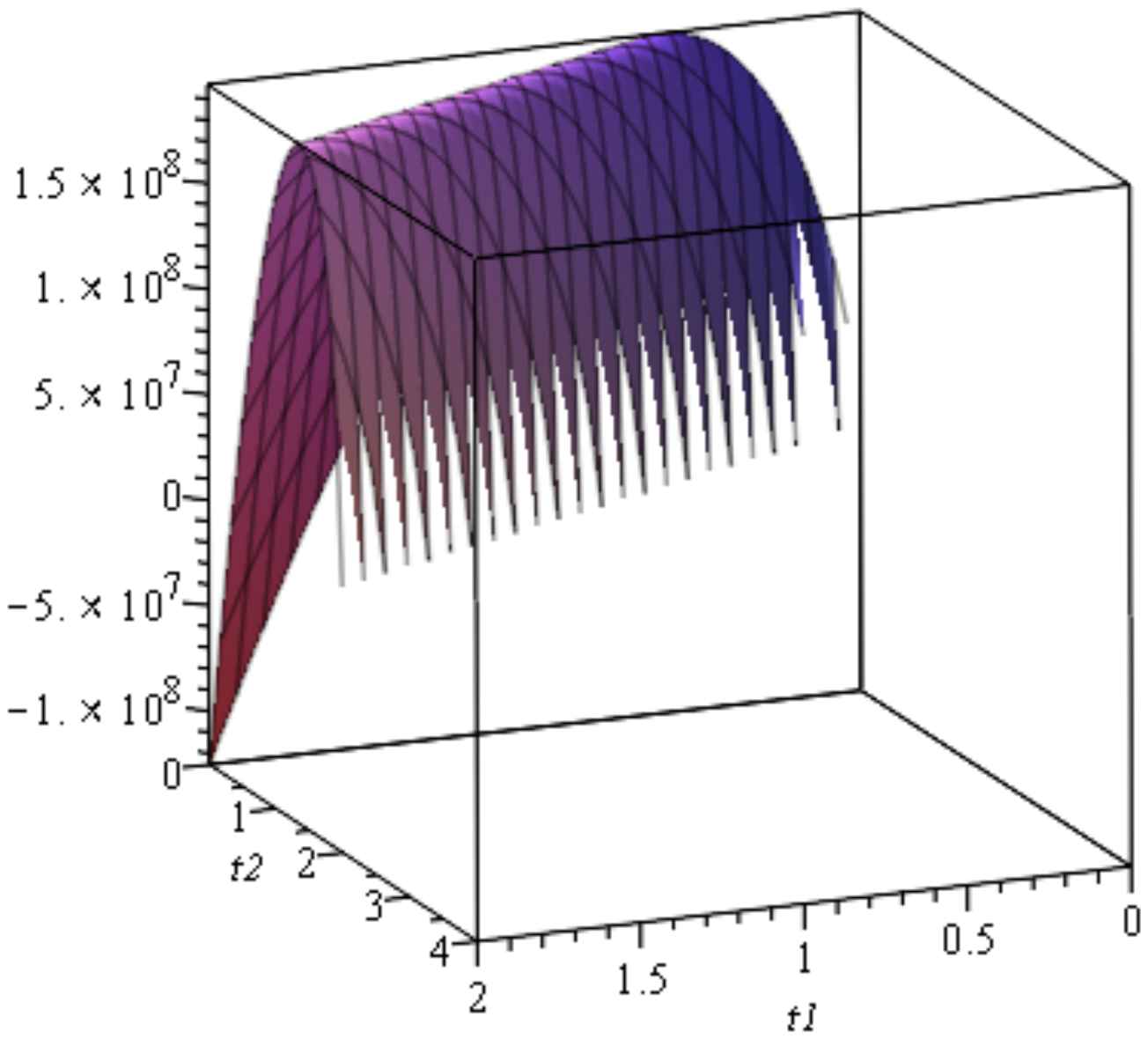}
		\caption{The two point correlation function $q_2(t_1,t_2)$ plotted as a function of the slab dimensions $\{ t_1, t_2 \}$ on $X$-axis and $Y$-axis respectively and the nature of the limiting $q_2$ along the $Z$-axis by considering it as the embedding function of the parametric configuration under fluctuations of $t_1$ and $t_2$ with the lattice volume $V=32$.} \label{fig7}
		\vspace*{-0.01cm}
	\end{figure}
\end{center}

\begin{center}
	\begin{figure}
		\hspace*{1.0cm}\vspace*{-7.6cm}
		\includegraphics[width=13.0cm,angle=0]{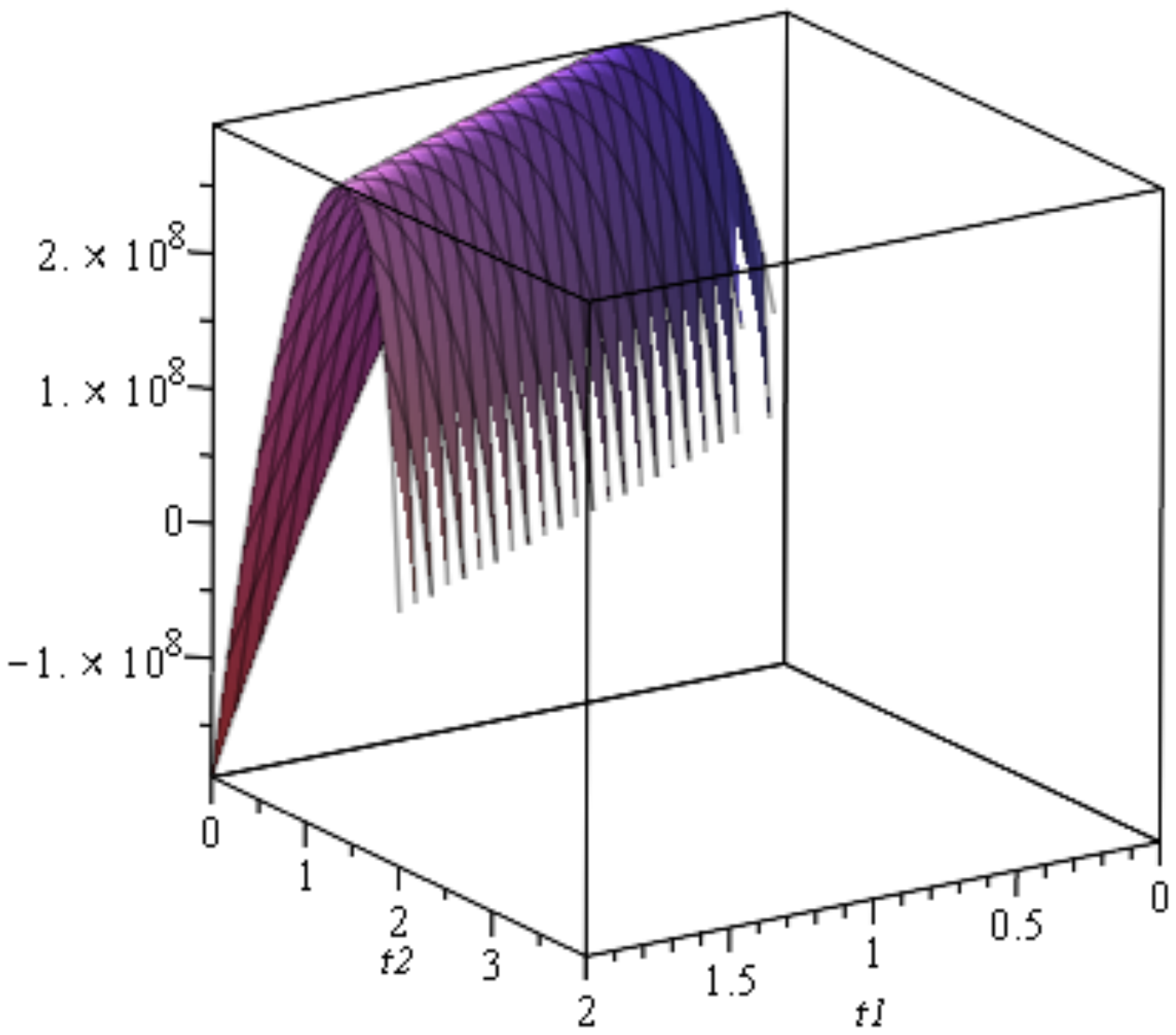}
		\caption{The two point correlation function $q_2(t_1,t_2)$ plotted as a function of the slab dimensions $\{ t_1, t_2 \}$ on $X$-axis and $Y$-axis respectively and the nature of the limiting $q_2$ along the $Z$-axis by considering it as the embedding function of the parametric configuration under fluctuations of $t_1$ and $t_2$ with the lattice volume $V=48$.} \label{fig7a}
		\vspace*{-0.01cm}
	\end{figure}
\end{center}

Hereby, we observe that $q_1(t_1,t_2)$ has a comb like behavior with respect to a slab dimension. For the lattice volume $V=48$, the behavior of the two point correlation function $q_1(t_1,t_2)$ is offered in Fig.(\ref{fig6a}). In this concern, Fig.(\ref{fig7}) illustrates the nature of the two point correlation function $q_2(t_1,t_2)$ at a subsequent lattice point as a function of the slab dimensions $\{ t_1, t_2 \}$ for the lattice volume $V=32$. The corresponding fluctuation behavior of   $q_2(t_1,t_2)$ for the lattice volume $V=32$ in shown in Fig.(\ref{fig7a}). In all the above cases, we notice that both the field theoretic correlations $q_1(t_1,t_2)$ and $q_2(t_1,t_2)$ have the same amplitude of fluctuations.


\begin{center}
	\begin{figure}
		\hspace*{1.0cm}\vspace*{-7.6cm}
		\includegraphics[width=13.0cm,angle=0]{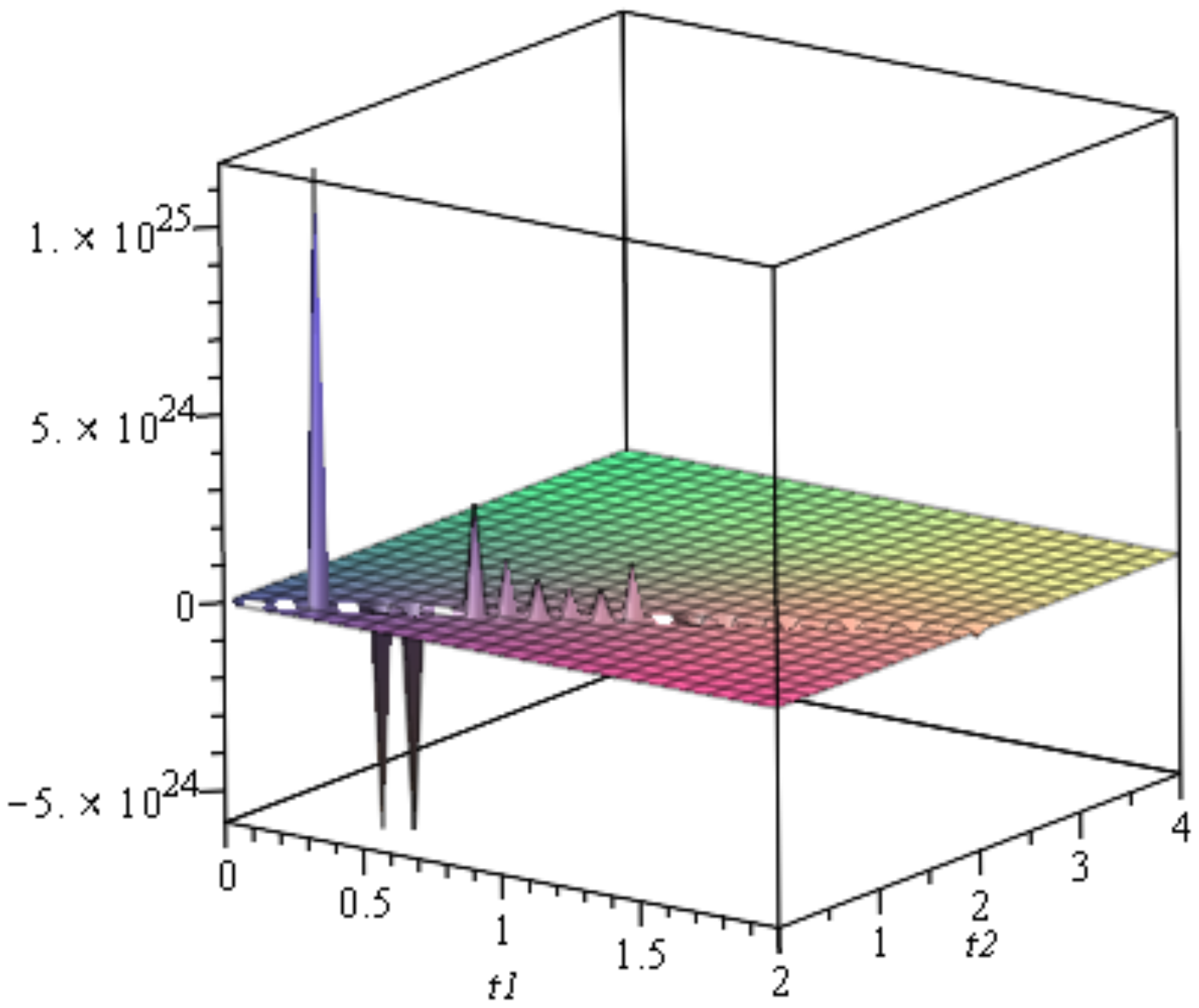}	
		\caption{The slab fluctuation capacity $q_{ii}(t_1,t_2)$ plotted as a function of the slab dimensions $\{ t_1, t_2 \}$ on $X$-axis and $Y$-axis respectively and $q_{ii}$ along the $Z$-axis by considering $q$ as the embedding function of the parametric configuration under fluctuations of $t_1$ and $t_2$ with the lattice volume $V=32$, where $i= 1, 2$.} \label{fig8}
		\vspace*{-0.01cm}
	\end{figure}
\end{center}
\begin{center}
	\begin{figure}
		\hspace*{1.0cm}\vspace*{-7.6cm}
		\includegraphics[width=13.0cm,angle=0]{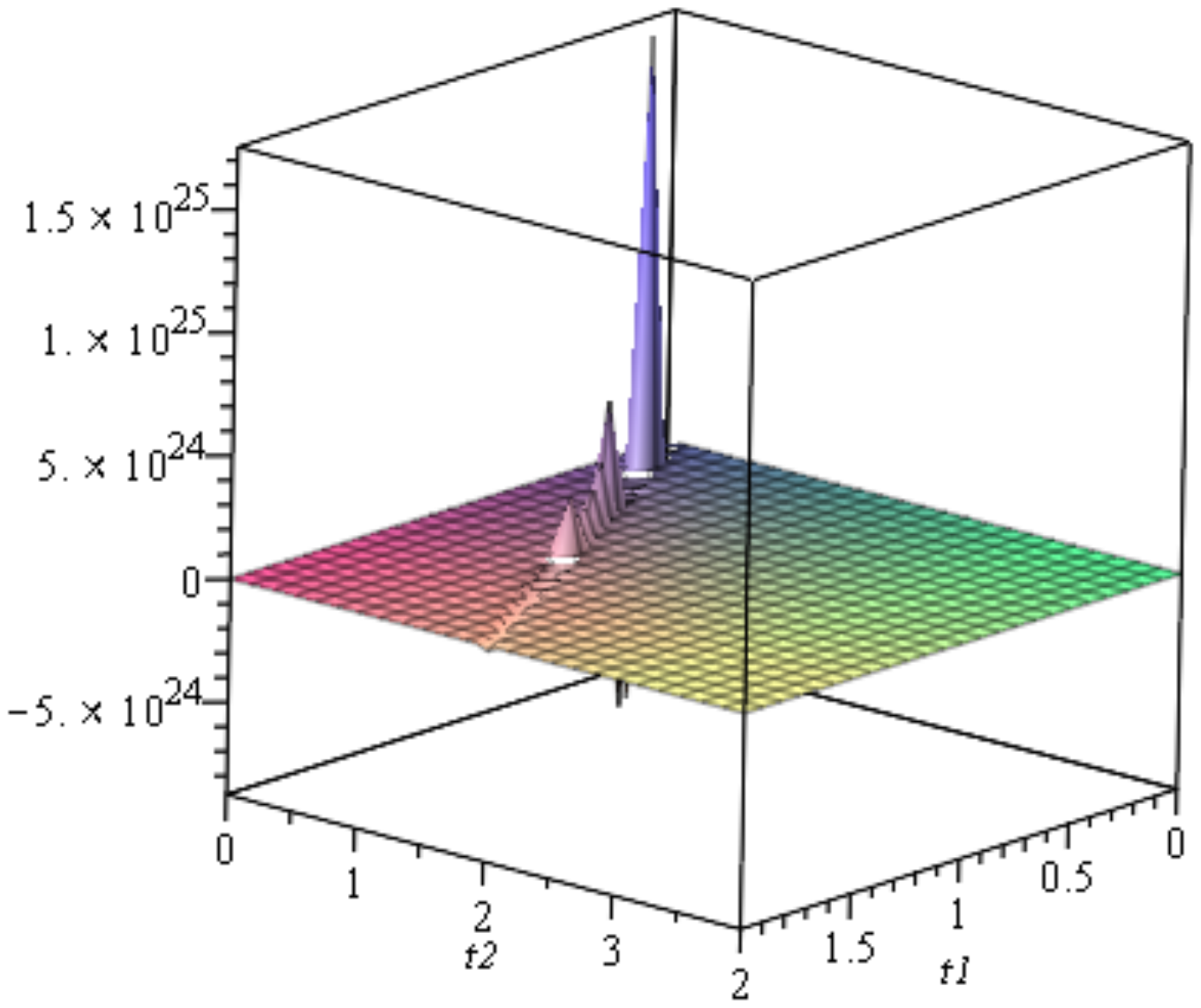}	
		\caption{The slab fluctuation capacity $q_{ii}(t_1,t_2)$ plotted as a function of the slab dimensions $\{ t_1, t_2 \}$ on $X$-axis and $Y$-axis respectively and $q_{ii}$ along the $Z$-axis by considering $q$ as the embedding function of the parametric configuration under fluctuations of $t_1$ and $t_2$ with the lattice volume $V=48$, where $i= 1, 2$.} \label{fig8a}
		\vspace*{-0.01cm}
	\end{figure}
\end{center}
%

%
%

In the sequel, we provide qualitative properties of the slab fluctuation capacities $\{q_{ii}(t_1,t_2)\ |\ i= 1,2\}$. Herewith, from Fig.(\ref{fig8}) we see that the slab fluctuation capacity $q_{ii}(t_1,t_2)$ plotted as a function of the slab dimensions $\{ t_1, t_2 \}$ on $X$-axis and $Y$-axis respectively and $q_{ii}$ along the $Z$-axis by considering $q$ as the embedding function of the fluctuation configuration has large positive and negative amplitudes of its order $\pm 10^{25}$ for the choice of the lattice volume $V=32$. Notice that there are further fluctuations of decreasing amplitudes situated along a straight line under an increment of the slab dimensions $ t_1$ and $t_2$. A similar fluctuation pattern is observed in Fig.(\ref{fig8a}) for the lattice volume $V=48$. In this case, we find a large amplitude of the same order when $ t_1$ and $t_2$ approach towards the origin, see Fig.(\ref{fig8a}). In this case, from Fig.(\ref{fig8a}) we equally observe that there are fluctuations of decreasing amplitudes along a straight line as we increase the slab dimensions $t_1$ and $t_2$ on a simulated lattice.

\begin{center}
	\begin{figure}
		\hspace*{1.0cm}\vspace*{-7.6cm}
		\includegraphics[width=13.0cm,angle=0]{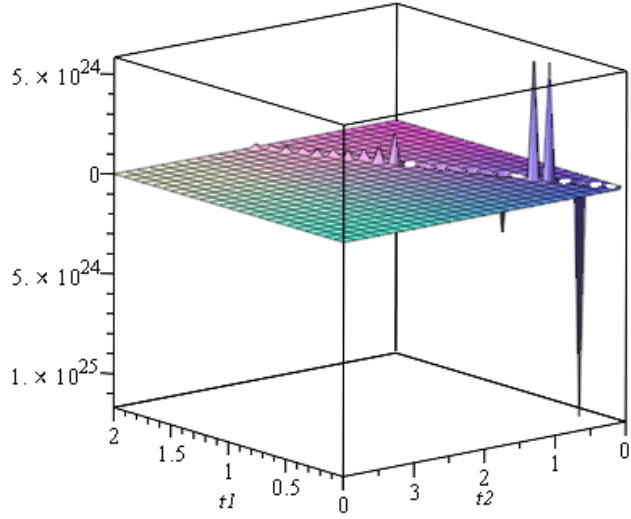}
		\caption{The slab dimension correlation $q_{12}(t_1,t_2)$ plotted as a function of the slab dimensions $\{ t_1, t_2 \}$ on $X$-axis and $Y$-axis respectively and $q_{12}$ along the $Z$-axis by considering $q$ as the embedding function of the parametric configuration under fluctuations of $t_1$ and $t_2$ with the lattice volume $V=32$.} \label{fig9}
		\vspace*{-0.01cm}
	\end{figure}
\end{center}
\begin{center}
	\begin{figure}
		\hspace*{1.0cm}\vspace*{-7.6cm}
		\includegraphics[width=13.0cm,angle=0]{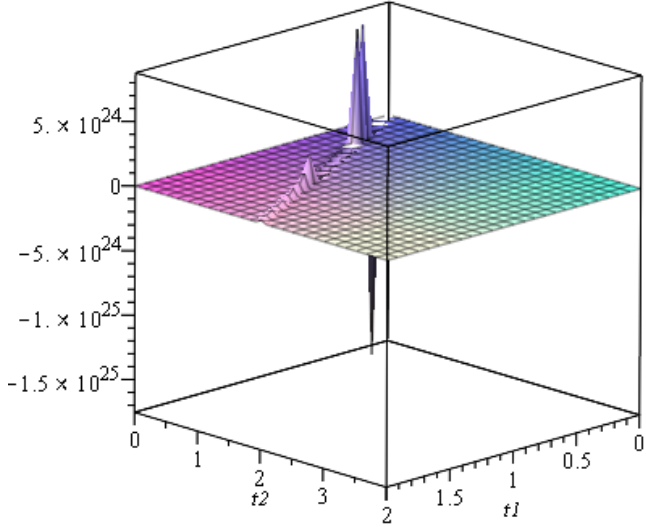}
		\caption{The slab dimension correlation $q_{12}(t_1,t_2)$ plotted as a function of the slab dimensions $\{ t_1, t_2 \}$ on $X$-axis and $Y$-axis respectively and $q_{12}$ along the $Z$-axis by considering $q$ as the embedding function of the parametric configuration under fluctuations of $t_1$ and $t_2$ with the lattice volume $V=48$.} \label{fig9a}
		\vspace*{-0.01cm}
	\end{figure}
\end{center}
%


In the light of above observations, the qualitative properties of the slab dimension correlation $q_{12}(t_1, t_2)$ is depicted in Figs.(\ref{fig9}, \ref{fig9a}) for the lattice volume $V=32$ and $V=48$ respectively. Namely, from Fig.(\ref{fig9}), we notice that the slab dimension correlation $q_{12}(t_1,t_2)$ (plotted against the slab dimensions $ t_1$, and $t_2$ on $X$-axis and $Y$-axis respectively and $q_{12}$ along the $Z$-axis by considering $q$ as the embedding function on the fluctuation surface) equally has large positive and negative amplitudes of fluctuations of their orders $\pm 10^{25}$ that is independent of the choice of the lattice volume, viz. it remains the same for $V=32$ and $V=48$. 

Furthermore, notice that there are fluctuations of decreasing amplitudes situated along a straight line under an increment of the slab dimensions $ t_1$ and $t_2$ in both the above cases of $V=32$ and $V=48$. The fluctuation pattern as observed in Fig.(\ref{fig9}) for $V=32$ is opposite and diagonally changed for the case of the lattice volume $V=48$. Finally, we find that there exists a large amplitude of the same order when $ t_1$ and $t_2$ approach towards the origin in either of the above cases, see Figs.(\ref{fig9}, \ref{fig9a}) on the above simulated lattices.

%
%


\subsection{Properties of Model Fluctuations}

In this subsection, we give physical implications of the slab sub-volume method of computing the topological susceptibility when its parameters, viz. the slab dimensions $\{t_1, t_2\}$ are allowed to fluctuate. For an arbitrary simulated lattice, the slab sub-volume method based fluctuation theory analysis has the following properties towards its validation:

\begin{itemize}
	\item It provides a good testing ground in comparison with the naive definition of the topological susceptibility $\chi_t^{slab}:= \langle Q^2(t_c) \rangle/V $ fluctuations based on the global topological charge. This is because the slop $\chi_t^{slab}$ is well consistent with the global topology, see \cite{Aoki, ALPHA} towards (2+1) QCD, chiral fermions and the determination of the strong coupling constant $\alpha_s$, whereby the associated fluctuations. 
	\item Specifically, our model  adequately represents the statistical and systematic errors by appropriately choosing the slab reference thicknesses $\{ t_1, t_2 \}$ of an ensemble of slab sub-volumes of an arbitrary simulated lattice. This has a linear suppression near the limit of the chiral extrapolation having a strong bias in the global topological sector.
	\item It shows a good agreement with the ChPT prediction \cite{2,3,4} by performing a two parameter fit with an estimation of the systematic effects that are negligibly small in comparison to the statistical and systematic errors.
	\item Namely, at the lightest up/ down quark mass, it follows that  $\chi_t^{slab}$ has a reasonable slop with respect to a chosen time cutoff of the slab.
	\item On the other hand, for the heavier quark masses, we find that the slab fluctuations show a nonzero scalar curvature that arises as an effect of the bias of the global topological charge $ \langle Q_{lat} \rangle $ over an arbitrary lattice $\mathbb{L}$.
	\item This supports the fact \cite{1} that the heavier pion mass ensembles show a longer auto-correlation than that of the smaller mass ensembles, whereby the respective sample yields a larger deviation in $ \langle Q_{lat} \rangle $ from zero. 
    \item It has a well controlled systematics due to finite volume effects and it has a consistent behavior with ensembles of heavier quark mass. This effectively determines the chiral condensate as a lower quark mass sample has good extrapolation behavior for the systematic error with the data in both the chiral and continuum limits.
    \item Herewith, this estimates systematic errors arising due to a non-linear behavior of different reference times. Namely, it possesses a large deviation as a systematic error with a double exponential structure of the undermining auto-correlation function as proposed by the ALPHA \cite{ALPHA} and JLQCD \cite{JLQCD} collaborations towards the determination of high energy systems and QCD coupling scales, also see \cite{1, Aoki}. 	
	\item It is free from the spatial divergences at equal spatial coordinates once the Yang-Mills gradient flow is performed, whereby two slabs of the same thickness are equivalent due to the transnational invariance of the model. 
	\item In addition, our analysis does not stop here, but it incorporates other systematic effects such as sizable finite volume fluctuations as the  associated  Yang-Mills gradient flow time dependence has a negligible effect on the system. Hereby, our fluctuation theory investigation is well-suited for numerical simulations and large scale data analysis programs in the light of QCD fundamentals and computational science.
\end{itemize}

In our analysis, the long range auto-correlations and associated slab biasing are optimized by requiring the vanishing of the scalar curvature of a nondegenerate surface of fluctuations of the slab dimensions $\{t_1, t_2\}$ as the model parameters. Namely, in this paper, we examine local and global stability structures and phase transition curves that arises through an ensemble of slab sub-volumes of an arbitrary simulated lattice.

\section{Conclusion}
In this paper, we study topological susceptibility fluctuations in (2+1) flavor QCD in order to explore stability properties in the light of high energy accelerator science. We compute topological susceptibility fluctuations and underlying correlation properties of an ensemble of (2+1) flavor QCD configurations with chiral fermions. We explore fluctuation theory analysis of an ensemble of dynamical M\"{o}bius domain wall fermions and compute the long rang correlation length by invoking the topological susceptibility as the model embedding function.

In particular, we have investigated quark mass dependence of the topological susceptibility for a variable mass scale that corresponds to renormalized pion masses. Physically, such a configuration enables in understanding of the structures of small quantum fluctuations. With the fact that the mass scale $M$ related to cutoff is kept arbitrary and there always exit statistical and systematic errors, the quark mass $m$ and the physical mass scale $M$ fluctuate independently. Hereby, given an arbitrary QCD configuration, we have explored the fluctuation theory analysis towards the optimal understanding of QCD configurations in the space of the mass of up/ down quarks $m$, and the physical mass scale $M$ corresponding to renormalized pion masses. Notice that the parameter $M$ is not the pion bare mass, but it is taken as the physical mass scale, see \cite{Aoki} in the light of chiral fermions and associated slab sub-volume method. In this paper, we have intrinsic geometrically addressed the issue of topological susceptibility fluctuations by realizing it as the embedding from the surface of the quark mass $m$ and the physical mass scale $M$ that corresponds to renormalized mass of pions to the set of real numbers. 

To be precise, by considering the topological susceptibility as the model embedding function, we compute parametric fluctuation properties of QCD vacuum in the light of phenomenological models, viz. the ChPT formulation and slab sub-volume method \cite{Aoki}. Following the fact that the quark mass dependence of the topological susceptibility arises from the quark sea, e.g., the quantum fluctuations that are suppressed by some orders of $\hbar$. Hereby, in the light of the ChPT formulation \cite{Aoki}, it is worth examining whether the physical mass scale $M$ (that corresponds to the renormalized mass of pions) is independent from the quark masses $m_u$ and $m_d$ or it depends on $m_u$ and $m_d$, or  on other parameters, for example, the chosen renormalization (e.g., $\overline{MS}$ scheme \cite{ms}), or the choice of a gauge (e.g., Symanzik gauge \cite{22,23,24}), see \cite{ow} towards composite Higgs models, and related constraints. In this consideration, our analysis offer the intrinsic geometric role of QCD towards the optimal designing of ATLAS detectors. Notice that the parameter $M$ is the physical value of the mass scale that is subject to vary in order to achieve renormalized pion masses. Hereby, the scale dependence, e.g., asymptotic freedom \cite{af}, QCD running coupling, parameter space transformations, Riemannian geometric investigations and associated optimal determination of $\alpha_s$ are in the future scope of this research.

Our analysis concentrates on fluctuations of topological susceptibility in an ensemble of varying slab sub-volumes of an arbitrary simulated lattice. Hereby, we offer explicit expressions of the fluctuation quantities including the local and global correlations, stability structure through the symmetric Hessian matrix as the metric tensor and associated global correlation length as the square root of the corresponding scalar curvature of the fluctuation surface. 
The geometric invariants emerging from the global topological susceptibility as the embedding map agree well with that of the limiting chiral perturbation theory for an ensemble of slab sub-volumes of a simulated lattice in the framework of fluctuation theory. This gives an intrinsic geometric understanding of vacuum fluctuations undermining a chiral condensate. In this concern, considering the quark mass $m$ and the mass scale $M$ corresponding to renormalized pion masses as a pair of statistical parameters, we find the respective surface of fluctuations corresponds to a non-interacting statistical configuration. 

On the other hand, the fluctuation surface that is formed by variations of the slab dimensions $\{t_1, t_2\}$ corresponds to an interacting statistical system except for a few values of the reference slab difference $t:=t_1-t_2$. 
Following the same,  we have explicitly computed the concerning correlation area as the ratio of two finite degree polynomials in $t$. Hereby, we have classified the stability regions and associated phase transition curves of the chosen observation sample. 
In the sequel, we have examined the intrinsic geometric equivalence of the slab sub-volume method of computing the topological susceptibility fluctuations with its counterpart as the ChPT formulation. Following the fact that ChPT configurations yields a non-interacting statistical basis in the space of quark mass $m$ and the mass scale $M$ corresponding to renormalized pion masses, we find that the slab sub-volume scheme yields an ill-defined statistical system in the infinitesimal limit of its parameters $t_1$ and $t_2$. Namely, given the topological susceptibility as the model embedding function, it is worth mentioning that the ChPT ensemble is always non-interacting in the space of $\{m, M\}$. However, the finite slab sub-volume ensemble generically corresponds to an interacting statistical basis under fluctuations of $\{t_1, t_2\}$.     

Using the dynamical M\"obius domain wall fermions, the slab sub-volume ensemble of a simulated lattice enables an intrinsic geometric examination of the topological susceptibility fluctuations of an arbitrary (2+1) flavor QCD where the quark mass dependent fluctuation quantities are consistent with the ChPT observations. Here, the fluctuations in the topological susceptibility comes from the statistical uncertainty in the space of the mass of the up/ down quarks, and the mass scale $M$ that corresponds to renormalized pion masses at each simulation point of a chosen lattice. This is because of the fact that our consideration is well-suited to examine effects coming from the topological freezing and systematics of the limiting chiral and continuum configurations, and that of the chiral condensate through the Dirac spectrum \cite{31} with an estimation of the next to the leading order effects that include the renormalized pion masses and lattice QCD based renormalized group invariants. Furthermore, it would be interesting to explore the $\theta$ dependence of ChPT formula and associated chiral fermion based coupling of the QCD to axion couplings to nucleons.  

Issues worth examining further towards a refined understanding of topological susceptibility fluctuations include it's concavity, relative entropy, associated quantum measurements in the framework of the density matrix, statistical correlations and their relations to quantum mechanics, viz. the entanglement of an ensemble of states in a given Hilbert space, and the monogamy of entanglement, for an overview, see \cite{Witten} in relation to the information theory and \cite{witteng} for entanglements in quantum field theory, and their relations to the complex condensed matter systems \cite{tcb} and embedding geometries \cite{bntef}. Hereby, by invoking the role of joint probability distribution of observations as fluctuations of the ChPT parameters $\{m, M\}$, or that of the slab sub-volume parameters $\{t_1, t_2\}$ of an arbitrary simulated lattice, it is worth exploring implications of our results towards Riemannian geometric studies, statistical correlations, simulation techniques, generalized measurements, probability distributions, and quantum channel measurements. Such issues we leave open for future research investigations.     

\section*{Acknowledgment}
	
We would like to thank  the Yukawa Institute for Theoretical Physics at Kyoto University.
Discussions during the workshop YITP-T-18-04 ``New Frontiers in String Theory 2018" were useful towards the completion of this paper. 
It is the matter of the pleasure to thank Prof. S. Aoki for an enlightening discussion on the slab sub-volume fluctuations of an arbitrary simulated lattice, and Prof. E. Witten for his suggestions on the physical mass scales of QCD.

\end{document}